\documentclass[pra,twocolumn,amsmath,amssymb,floatfix,reprint,footinbib,superscriptaddress,longbibliography,showkeys]{revtex4-2}
\usepackage{picinpar,graphicx}
\usepackage{braket}
\usepackage{amsmath}
\usepackage{graphicx}
\usepackage{dcolumn}
\usepackage{graphicx}
\usepackage{mathrsfs}
\usepackage{mdwlist}
\usepackage{subfigure}
\usepackage{booktabs}
\usepackage{amsmath}
\usepackage{dsfont}
\usepackage{amstext}
\usepackage{amssymb}
\usepackage{amsbsy}
\usepackage{bbm}
\usepackage{amsthm}
\usepackage{graphicx}
\usepackage{textcomp}
\usepackage{multirow}
\usepackage{color}
\usepackage[colorlinks,citecolor=blue]{hyperref}
\definecolor{Dgreen}{RGB}{0, 100, 0}
\usepackage{url}
\usepackage[colorlinks]{hyperref}
\hypersetup{%
	plainpages=true,
	breaklinks=true,  
	hypertexnames=false,  
	pageanchor=true,
	colorlinks=true,
	linkcolor={blue},
	citecolor={red},
	urlcolor={blue},
	anchorcolor={black}
}

\begin{document}
	
	\title{Generating three-photon Rabi oscillations without a large-detuning condition}
	\author{Ke-Xiong Yan}
	\affiliation{Department of Physics, Fuzhou University, Fuzhou, 350116, China}
	\affiliation{Fujian Key Laboratory of Quantum Information and Quantum Optics, Fuzhou University, Fuzhou 350116, China}
	
	\author{Yuan Qiu}
	\affiliation{Department of Physics, Fuzhou University, Fuzhou, 350116, China}
	\affiliation{Fujian Key Laboratory of Quantum Information and Quantum Optics, Fuzhou University, Fuzhou 350116, China}
	
	\author{Yang Xiao}
	\affiliation{Department of Physics, Fuzhou University, Fuzhou, 350116, China}
	\affiliation{Fujian Key Laboratory of Quantum Information and Quantum Optics, Fuzhou University, Fuzhou 350116, China}
	
	\author{Ye-Hong Chen}\thanks{yehong.chen@fzu.edu.cn}
	\affiliation{Department of Physics, Fuzhou University, Fuzhou, 350116, China}
	\affiliation{Fujian Key Laboratory of Quantum Information and Quantum Optics, Fuzhou University, Fuzhou 350116, China}
	\affiliation{Theoretical Quantum Physics Laboratory, RIKEN Cluster for Pioneering Research, Wako-shi, Saitama 351-0198, Japan}
	
	\author{Yan Xia}\thanks{xia-208@163.com}
	\affiliation{Department of Physics, Fuzhou University, Fuzhou, 350116, China}
	\affiliation{Fujian Key Laboratory of Quantum Information and Quantum Optics, Fuzhou University, Fuzhou 350116, China}

	\date{\today}
	
	\begin{abstract}
		It is well known that in the quantum Rabi model, a three-photon resonance occurs when the cavity field bare frequency is about 1/3 of the atomic transition frequency. In this manuscript, we show that the resonance can also be generated in the absence of the ``1/3 condition" by employing an artificial atom with tunable transition frequency. To realize the protocol, the modulation frequency should be comparable to the cavity frequency in order to induce a counter-rotating interaction in the effective Hamiltonian. In this way, three-photon Rabi oscillations can be observed in a small-detuning regime, thus avoiding the excitation of high-energy states. We derive an effective Hamiltonian (equivalent to the anisotropic Rabi model Hamiltonian) to determine the magnitude of the energy splitting and the resonance position. Numerical simulations results show that the protocol not only generates a three-photon resonance, but also has a detectable output photon flux. We hope the protocol can be exploited for the realization of Fock-state sources and the generation of multiparticle entanglement.
	\end{abstract}
	
	\maketitle

	\section{INTRODUCTION}
	The quantum Rabi model (QRM)~\cite{PhysRev.49.324,4PhysRev.51.652,1PRL76.1800,PhysRevLett.107.100401,ashhab_rabi_2006}, describing the interaction of a two-level quantum system and a bosonic mode, is one of the fundamental physical models.
	Despite its simplicity, the QRM exhibits a rich set of physical properties. It has been used to describe the dynamics of various physical setups~\cite{kwek2013strong}, ranging from quantum optics~\cite{2WOS:001142919400001,3WOS:001137118100005} to condensed matter physics~\cite{PhysRevB.108.174306,PhysRevLett.128.160504}. The full Hamiltonian of the QRM includes the counter-rotating (CR) terms, the presence of which leads to non-conservation of the system excitation number~\cite{PhysRevA.95.063849,4WOS:000542185800009,5WOS:000470901400001,PhysRevA.105.023720,PhysRevA.102.033716,PhysRevLett.122.030402}.
	This excitation-number-nonconserving process produces virtual photons, enabling high-order atom-field resonant transitions~\cite{-PhysRevA.95.063849}. Hence, one can realize analogs of many nonlinear-optics effects~\cite{PhysRevA.95.063849,Chen2024,PhysRevA.84.033823,PhysRevA.85.023809}, such as various frequency-conversion processes~\cite{Akbari2016,PhysRevA.106.053503}, parametric amplification~\cite{PhysRevLett.85.2308}, multiphoton absorption~\cite{Alam2017}, Kerr effect~\cite{PhysRevLett.89.047401}, and other nonlinear processes~\cite{8WOS:000366730900005,PhysRevLett.117.043601,PhysRevA.88.033822}. 
	
	In general, the high-order resonant transition between system states requires a strong light-matter coupling to be noticeable~\cite{4WOS:000542185800009, 5WOS:000470901400001,5Ridolfo2012,5Felicetti2015,5Cao2011,6WOS:000286732400017,13Garziano2016,16Ma2015,15Felicetti2018,8WOS:000366730900005,10WOS:001112616000002}. When the atomic transition frequency and the cavity frequency satisfy a certain relation, this transition is also to be visible even if the coupling strength is weak.
	For instance~\cite{16Ma2015,8WOS:000366730900005}, a three-photon-resonance transition can be realized in a weak coupling regime as long as the frequency of the bosonic cavity is close to one third of the atomic transition frequency (i.e.~``1/3 condition") through a virtual excitation process. This is an anomalous Rabi oscillation induced by the higher-order processes, in which three photons are co-emitted by the excited atom into a resonator and vice versa. Such a high-order resonance process can be used to realize Fock-state sources~\cite{PhysRevLett.88.143601,Shi2014} and to generate multiparticle entanglement~\cite{PhysRevA.103.032402}.
	
	However, the ``1/3 condition"~\cite{16Ma2015,8WOS:000366730900005} may lead to problems in experimental observation of the three-photon resonance. With such a large detuning, the anharmonicity of the artificial atom is too small (compared to detuning) to avoid excitation to the higher energy states. Addressing this problem, we propose a protocol to generate the three-photon resonance without such a large detuning. Specifically, by introducing a modulation to the energy separation of the artificial atom~\cite{17Huang2017,18Silveri2017,19Liao2015,20Liao2016,21Beaudoin2012,22Zhang2024,23Nourmandipour2023,PhysRevA.75.063414}, dynamics of the system can be described by an effective anisotropic Rabi Hamiltonian. By tuning the modulation frequency, the effective cavity frequency and the effective atomic transition frequency can satisfy the ``1/3 condition" resulting in the three-photon resonance. The calculations indicate that under such a longitudinal drive~\cite{PhysRevLett.131.113601}, the artificial atom does not jump to the third excited state or higher energy states, even if the energy level anharmonicity is much smaller than the atom transition frequency. In contrast to the protocol, the scheme~\cite{8WOS:000366730900005} using the large-dutuning condition fails to generate three-photon Rabi oscillations in superconducting circuits when considering the influence of the energy level anharmonicity.
	
	The rest of the paper is organized as follows: In Sec.~\ref{s2}, we introduce the model and the Hamiltonian. In Sec.~\ref{s3}, the evidence of the three-photon resonance in an effective anisotropic Rabi model is shown by analysing the energy spectrum of the system. At the same time, we also give theoretical formulas to calculate the resonant frequency and energy splitting. In Sec.~\ref{s5}, we calculate the output photon flux rate of the system in the presence and absence of frequency modulation. In Sec.~\ref{s4}, the influence of the high energy levels of the artificial atom on the three-photon resonance is discussed. In Sec.~\ref{s6}, the experimental parameters are provided. At last, the work is summarized in Sec.~\ref{s7}. 
	\section{MODEL} \label{s2}
\begin{figure}
		\centering
		\includegraphics[height=6cm,width=7.8cm]{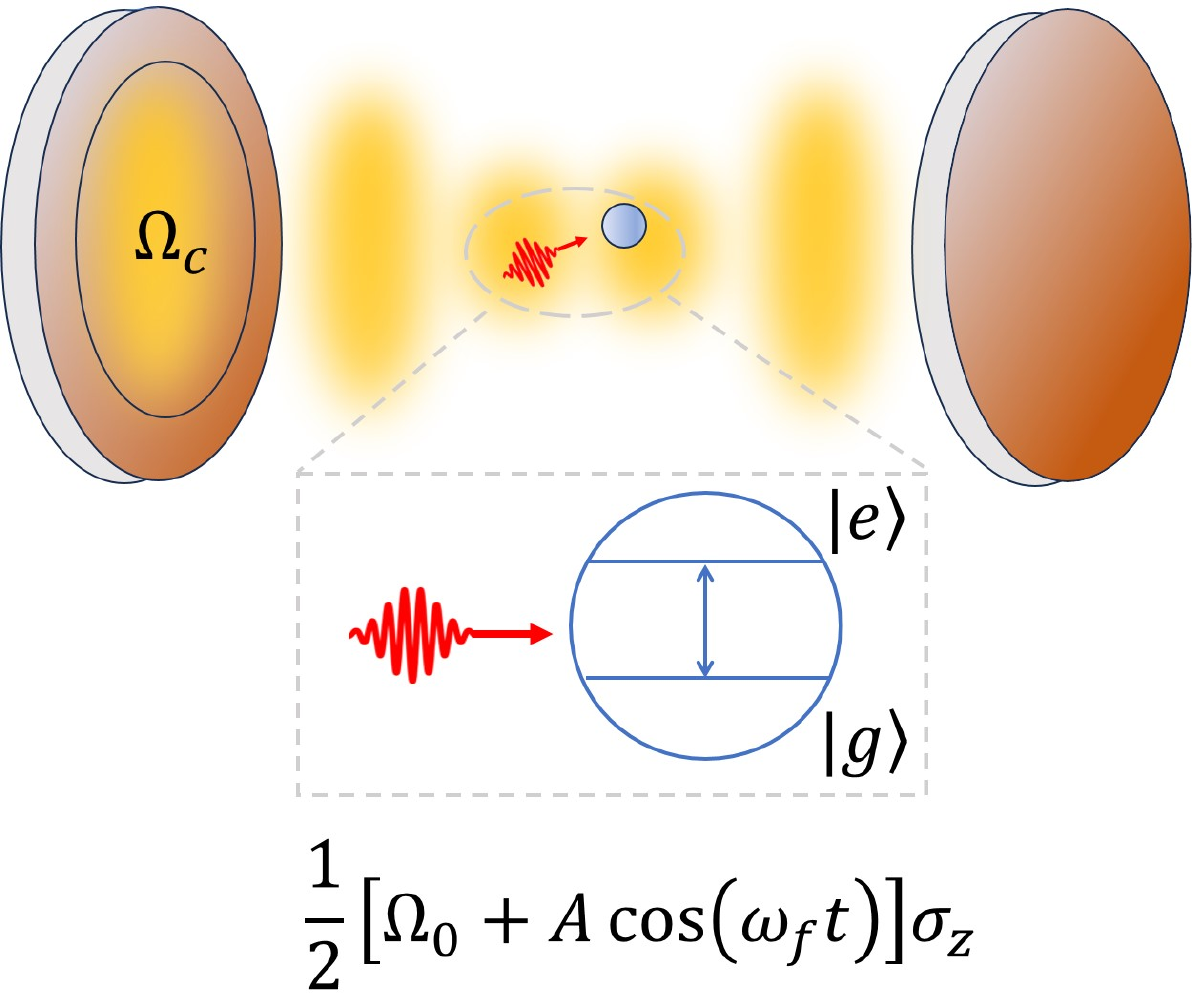}
		\caption{Schematic of the system with an artificial atom coupled to a single-mode cavity. The transition frequency of the artificial atom is $\Omega_{0}$ and the frequency of the cavity mode is $\Omega_{c}$. An applied driving field modulates the atomic transition frequency. The parameters $A$ and $\omega_{f}$ are respectively the modulation amplitude and the modulation frequency.}
		\label{F1}
	\end{figure}

	The QRM we consider (see Fig.~\ref{F1}) consists of an artificial atom with transition frequency $\Omega_{0}$ and a cavity with resonant frequency $\Omega_{c}$, where the energy levels  separation of the atom are modulated by an external flux bias. The Hamiltonian ($\hbar=1$) of this model is given by\\
	\begin{equation}
		\begin{aligned}
			H_{1}(t)&=H_{R}+H_{M}(t), \label{eq1}
		\end{aligned}
	\end{equation}
	where
	\begin{equation}
		H_{R}=\Omega_{c}a^{\dagger}a+\frac{\Omega_{0}}{2}\sigma_z+\lambda(a^{\dagger}+a)\sigma_{x}
	\end{equation}
	is the Hamiltonian of the QRM, and\\
	\begin{equation}
		H_{M}(t)=\frac{A\cos(\omega_{f}t)}{2}\sigma_z
	\end{equation}
	describes a sinusoidal modulation applied to the artificial atom. Here, $a^{\dagger}$ and $a$ are the creation and annihilation operators, respectively. The coupling strength between the cavity and the artificial atom is $\lambda$. The Pauli matrices are defined as $\sigma_z=\ket{e}\bra{e}-\ket{g}\bra{g}$ and $\sigma_{x}=\ket{e}\bra{g}+\ket{g}\bra{e}$, where the excited state and the ground state of the atom are $\ket{e}$ and $\ket{g}$, respectively. The modulation amplitude and frequency are $A$ and $\omega_{f}$, respectively. 

    To study the dynamics of the model, we can transform the Hamiltonian $H_{R}$ to the rotating frame defined by 
    \begin{equation}
	U_{1}(t)={\rm exp}\left\{-i\Omega_{c}ta^{\dagger}a-i[\frac{\Omega_{0}}{2}t+\frac{A}{2\omega_{f}} \sin(\omega_{f}t)]\sigma_z\right\}.
    \end{equation}  
    Thus, using the Jacobi-Anger expansion
    \begin{equation}
	e^{ix\sin(\omega_{f}t)}=\sum_{n=-\infty}^{+\infty}J_{n}(x)e^{in\omega_{f}t}, \label{eq5}
\end{equation}
 the Hamiltoian in Eq.~(\ref{eq1}) is written as
\begin{eqnarray}
	\nonumber
	H_{2}(t)&=&i\dot{U}^{\dagger}(t)U(t)+U^{\dagger}(t)H_{1}(t)U(t)\\ \nonumber
	&=&\sum_{n=-\infty }^{+\infty} \lambda J_{n}(x) a^{\dagger}\sigma_{+}e^{i(\Omega_{0}+\Omega_{c}+n\omega_{f})t}\\ 
	&&+\sum_{m=-\infty }^{+\infty} \lambda J_{m}(x) a\sigma_{+}e^{i(\Omega_{0}-\Omega_{c}+m\omega_{f})t}+\rm H.c.,\nonumber \\ \label{eq6}
\end{eqnarray}
where $x=A/\omega_{f}$. The transition operators of the atom are defined as $\sigma_{+}=\sigma_{-}^{\dagger}=\ket{e}\bra{g}$. $J_{n}(x)$ is the $n$th Bessel function of the first kind. For further discussion, we set $\delta=\Omega_{0}-\Omega_{c}$ and $\Delta=\Omega_{0}+\Omega_{c}-\omega_{f}$, as well as $J_{n}\equiv J_{n}(x)$, resulting in
\begin{eqnarray}
	H_{2}(t)&=&\lambda \left(J_{-1}a^{\dagger}\sigma_{+}e^{i\Delta t}+ J_{0}a\sigma_{+}e^{i\delta t}\right)\cr\cr
	&&+\sum_{n\neq-1} \lambda J_{n} a^{\dagger}\sigma_{+}e^{i[\Delta+(n+1)\omega_{f}]t}\cr\cr 
	&&+\sum_{m\neq0} \lambda J_{m} a\sigma_{+}e^{i(\delta+m\omega_{f})t}+\rm H.c..  \label{eq7} 
\end{eqnarray}
Under the conditions
\begin{equation}
	\omega_{f}\gg \left\{\delta,\Delta,\lambda \lvert J_{n}(x) \rvert\right\},
	\label{c8}
\end{equation}
we discard the fast oscillating terms in Eq.~$(\ref{eq7})$ by performing the rotating wave approximation (RWA). Then the Hamiltonian $H_{2}(t)$ in Eq.~$(\ref{eq7})$ can be simplified as
\begin{equation}
	\widetilde{H}_{2}(t)\approx \lambda_1a^{\dagger}\sigma_{+}e^{i\Delta t}
	+\lambda_{2}a\sigma_{+}e^{i\delta t}+\rm H.c.,
	\label{eq9}
\end{equation}
where $\lambda_1 = \lambda J_{-1}$ and $\lambda_2 = \lambda J_{0}$ are the effective coupling strengths.

The Hamiltonian $\widetilde{H}_{2}(t)$ in Eq.~$(\ref{eq9})$ is equivalent to the following anisotropic Rabi Hamiltonian~\cite{Xie2014}
\begin{equation}
	\begin{aligned}
		H_{aR}=\omega_{c}a^{\dagger}a+\frac{\omega_{0}}{2}\sigma_{z}
		+(\lambda_1a^{\dagger}\sigma_{+}+\lambda_2a\sigma_{+}+\rm H.c.), \label{eq10}
	\end{aligned}
\end{equation}
where $\omega_{c}=(\Delta-\delta)/2$ and $\omega_{0}=(\Delta+\delta)/2$ are the effective cavity frequency and effective atomic transition frequency, respectively. 

\section{THREE-PHOTON RESONANCE} \label{s3}
In this section, we give the parameter conditions for the three-photon resonance described by the anisotropic Rabi model Hamiltonian in Eq.~$(\ref{eq10})$, and derive the effective Hamiltonian which describes an effective three-photon coupling between $\ket{e,0}$ and $\ket{g,3}$.

\subsection{Avoided Crossing and Resonance}
\begin{figure}
	\centering
	\includegraphics[scale=0.4]{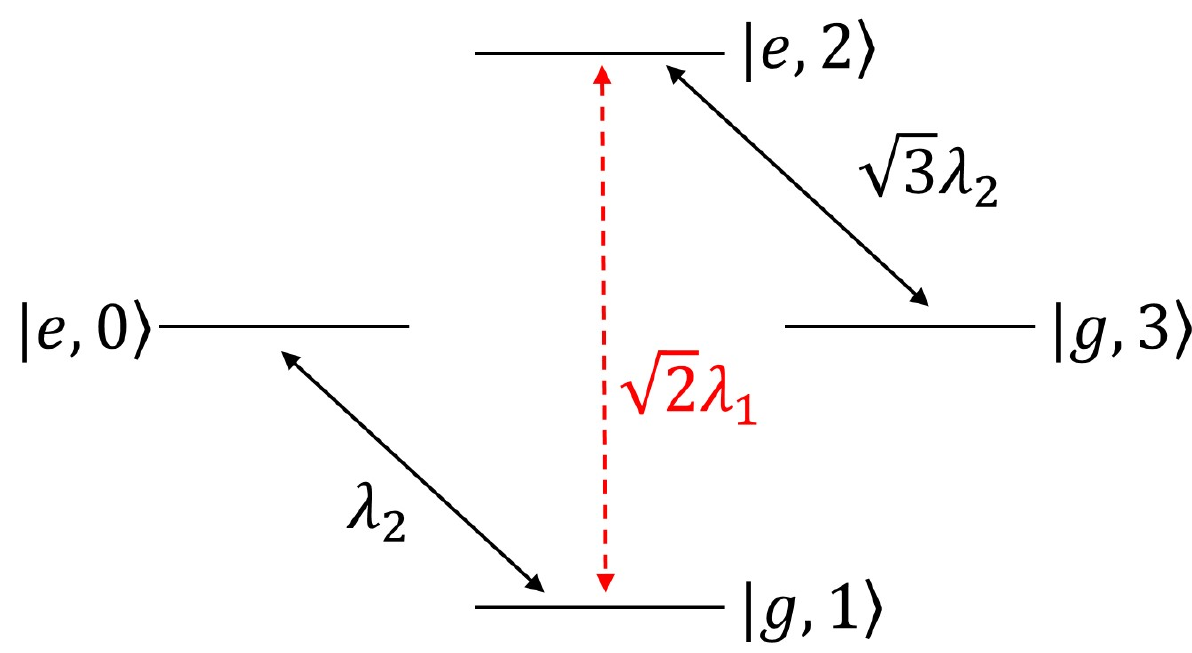}
	\caption{Sketch of the process giving the main contribution to the effective coupling between the bare states $\ket{e,0}$ and $\ket{g,3}$, via intermediate virtual transitions. The excitation-number-nonconserving processes are presented by the red arrow dashed line; $\lambda_{2}$, $\sqrt{2}\lambda_{1}$, and $\sqrt{3}\lambda_{2}$ are transition matrix elements.}
	\label{F2}
\end{figure}

\begin{figure}
	\centering
	\includegraphics[scale=0.43]{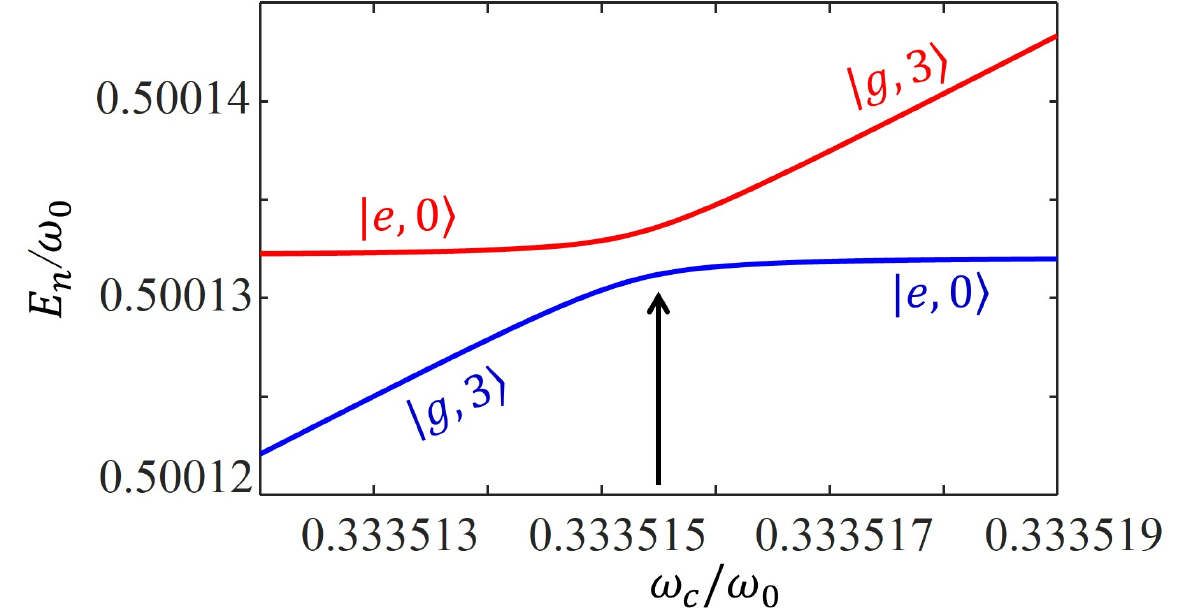}
	\caption{Eigenvalues $E_{3}/\omega_{0}$ and $E_{4}/\omega_{0}$ as a function of the ratio between $\omega_{c}$ and $\omega_{0}$ with $\lambda =0.01\omega_{0}$ and $x=0.5$. The vertical arrow indicates the frequency of the cavity corresponding to the occurrence of three-photon resonance, i.e. $\omega_{c}$$\approx$0.3335153$\omega_{0}$. The magnitude of the energy splitting is about $2.35\times10^{-6}\omega_{0}$. }
	\label{F3}
\end{figure}

\begin{figure}[t]
	\centering
	\includegraphics[scale=0.43]{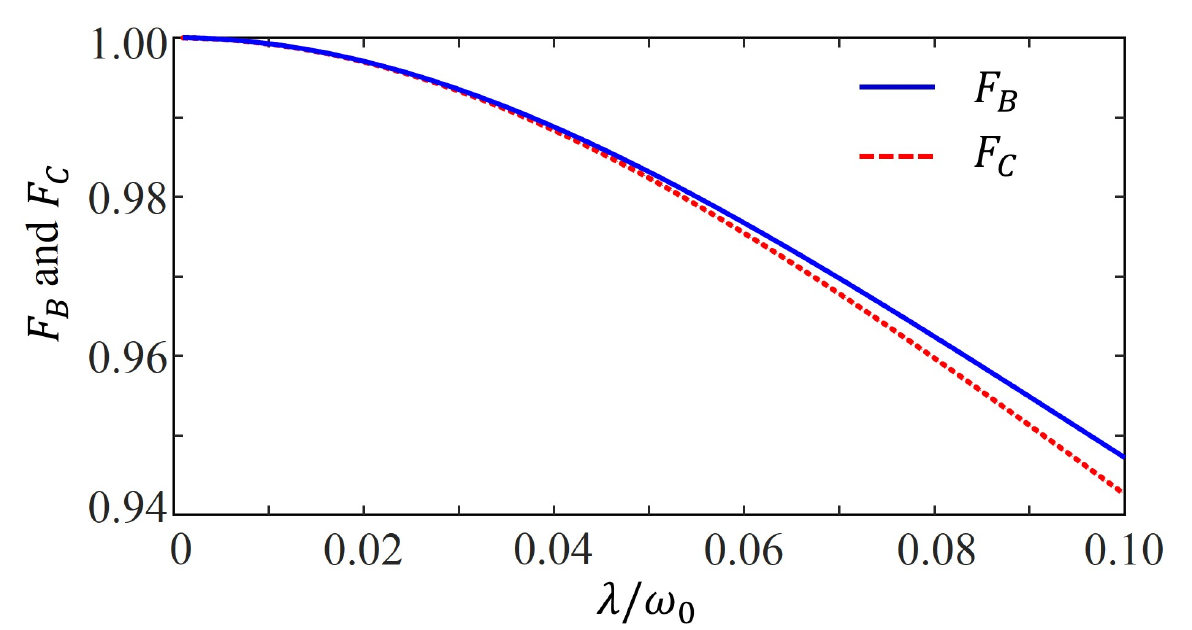}
	\caption{The overlaps $F_{B}$ (blue solid line) and $F_{C}$ (red dashed line) as functions of $\lambda/\omega_{0}$ at the avoided crossing point. The parameters are the same as in FIG.~$\ref{F3}$.}
	\label{F4}
\end{figure}

 We show that there is also a three-photon resonance in the anisotropic Rabi model. Figure~$\ref{F2}$ depicts the transition between states $\ket{e,0}$ and $\ket{g,3}$, and this transition is completely caused by excitation-number-nonconserving processes. In the parameter region of $x=0.5$ and $\lambda=0.01\omega_{0}\ll \delta$, we numerically calculate a portion of the energy spectrum for the Hamiltonian $H_{aR}$ as a function of $\omega_{c}$ and the results are plotting in Fig.~$\ref{F3}$. Here we focus on the third excited state and the fourth excited state of $H_{aR}$, i.e. $H_{aR}\ket{\psi_{n}}=E_{n}\ket{\psi_{n}}$ with $n=3,4$. In Fig.~$\ref{F3}$, we can find that the two energy levels exhibit an avoided crossing, which is a signature of the resonance between $\ket{e,0}$ and $\ket{g,3}$.  
 At this point, $\ket{\psi_{3}}$ and $\ket{\psi_{4}}$ can be well approximated by $\ket{B}=(\ket{e,0}+\ket{g,3})/\sqrt{2}$ and $\ket{C}=(\ket{e,0}-\ket{g,3})/\sqrt{2}$, respectively. By numerical calculations, the overlap, $F_{B}=\left| \langle B|\psi_{3}\rangle \right|$ and $F_{C}=\left| \langle C|\psi_{4}\rangle \right|$, at the resonance point as functions of $\lambda/\omega_{0}$ shows in Fig.~$\ref{F4}$. Both $F_{B}$ and $F_{C}$ are approximately equal to 1 for $\lambda<0.01\omega_{0}$. We note that the overlaps decrease as the coupling strength increases. That is because when the interaction strength is too strong, eigenstates of the Hamiltonian $H_{aR}$ are highly dressed so that cannot be described by $\ket{B}$ and $\ket{C}$.
 
\begin{figure}[t]
	\centering
	\subfigure{
		\label{F5(a)}
		\includegraphics[scale=0.41]{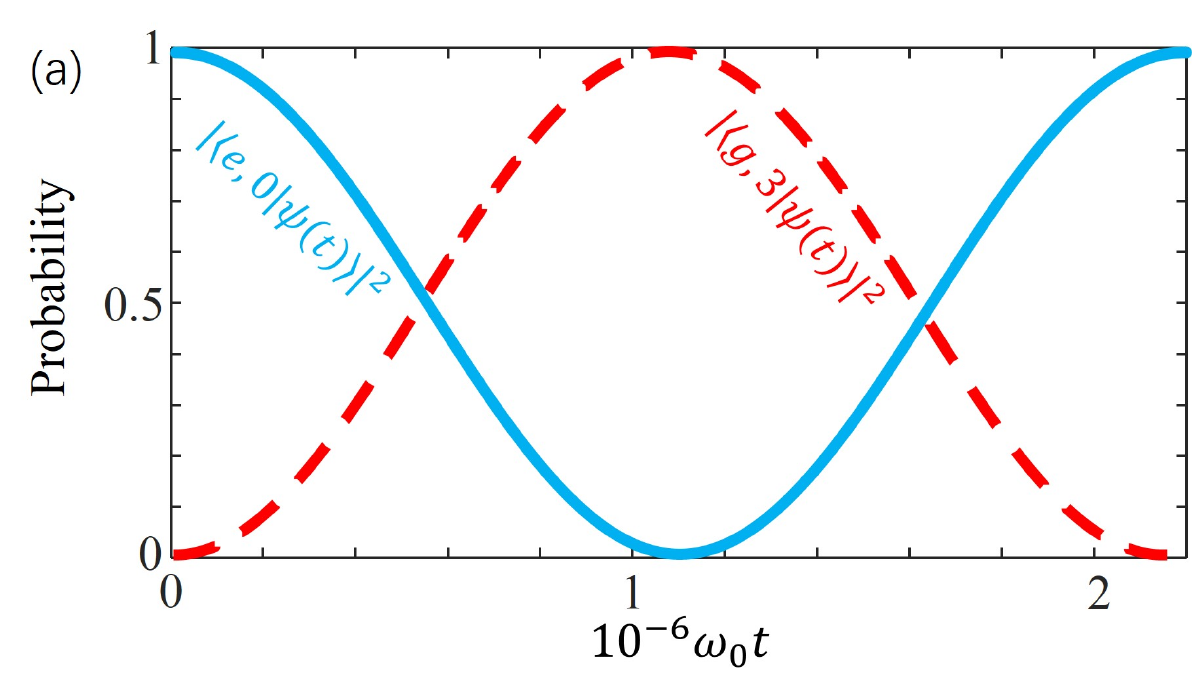}	
	}
    \subfigure{
    	\label{F5(b)}
    	\includegraphics[scale=0.41]{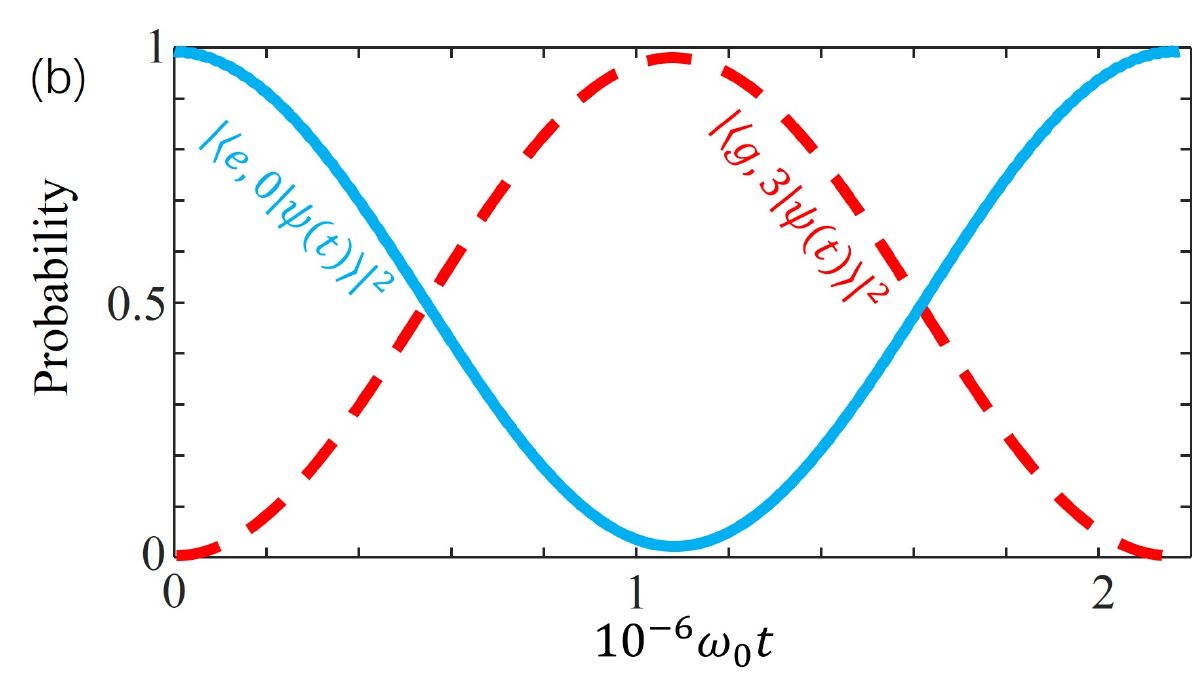}
    }
	\caption{Time dependence of the probabilities $|\langle e,0|\psi(t)\rangle|^{2}$ (blue solid curves) and $|\langle g,3|\psi(t)\rangle|^{2}$ (red dashed curves). The numerical simulation results of the dynamics governed by the Hamiltonian $H_{aR}$ in Eq.~(\ref{eq10}) and $H_{2}(t)$ in Eq.~(\ref{eq7}) are shown in (a) and (b), respectively. The highest probabilities of $\ket{g,3}$ in (a) and (b) are $99.85\%$ and $99.18\%$, respectively. The modulation frequency is $\omega_{f}=1.96\Omega_{0}$ and the modulation amplitude is $A=0.98\Omega_{0}$. All results are plotted as functions of $\omega_{0}t$ with $\lambda =0.01\omega_{0}$, $x=0.5$, $\Omega_{0}=100\omega_{0}$, and $\omega_{c}$$\approx$0.3335153$\omega_{0}$.}
	\label{F5}

\end{figure}

\begin{figure}
	\centering
	\includegraphics[scale=0.43]{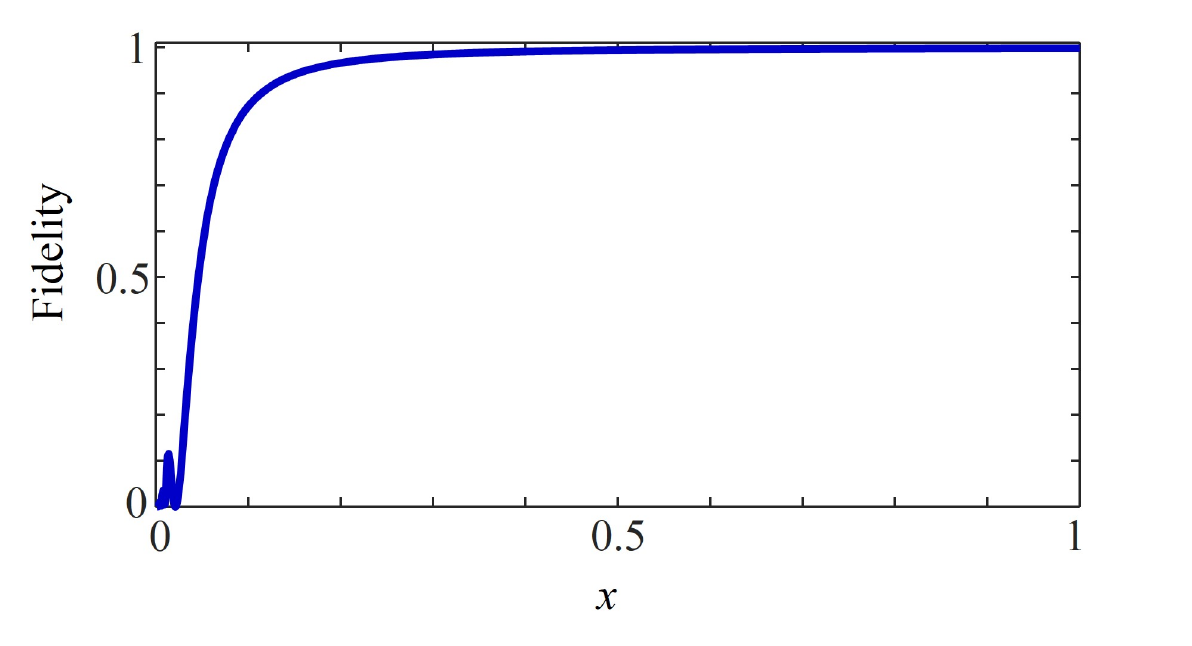}
	\caption{Fidelity of the state $\ket{g,3}$ as a function of $x$. The parameters are the same as in Fig.~$\ref{F5}$.}
	\label{F6}
\end{figure}

  Figure~\ref{F5} demonstrates that three-photon Rabi oscillations are generated in the parametric regime. It also shows that the approximate result agrees [Fig.~\ref{F5(a)}] well with the exact dynamics [Fig.~\ref{F5(b)}] when the conditions in Eq.~(\ref{c8}) are satisfied. We see that the probability of the three-photon state $\ket{g,3}$ can reach $99.18\%$. In the protocol, the parameter $x$ is the ratio of the modulation amplitude $A$ and the modulation frequency $\omega_{f}$, i.e. $x=A/\omega_{f}$. At a certain modulation frequency, the magnitude of the modulation amplitude $A$ affects the value of the parameter $x$, which in turn changes the fidelity of the three-photon state. To find out how the fidelity depends on the parameter $x$, we plot the fidelity of the state $\ket{g,3}$ at time $t \approx 1.1\times 10^{6}/\omega_{0}$ as a function of $x$ (see Fig.~\ref{F6}). We can find that when $x$ is taken as 0.5, the fidelity of $\ket{g,3}$ can reach nearby 1. In this case, the modulation amplitude $A$ is less than half of the modulation frequency $\omega_{f}$~\cite{17Huang2017}, ensuring that the present protocol is physically reasonable.
 
\begin{figure}
	\centering
	\includegraphics[scale=0.43]{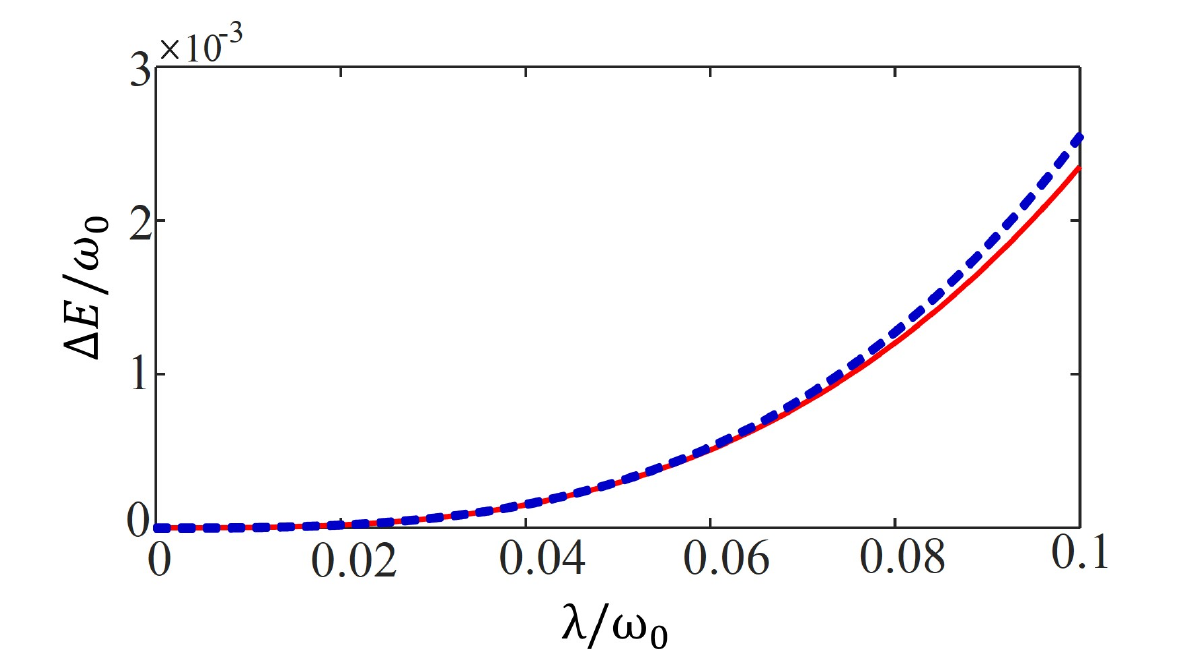}
	\caption{Comparison of the magnitudes of the energy splitting $\Delta E$ obtained analytically (red soild curve) and numerically (blue dashed curve) as a function of the interaction strength $\lambda/\omega_{0}$. Other parameters are the same as in Fig~~$\ref{F5}$.}
	\label{F7}
\end{figure} 
   
    \subsection{Effective Hamiltonian}
    We derive the effective Hamiltonian of $H_{aR}$ which determines the magnitude of energy splitting and the position of the resonance. To further understand how the three-photon resonance occur, we derive effective Hamiltonian using the high-order time-averaging method~\cite{PhysRevLett.85.2392,James2007,Shao2017}.
    
    For the Hamiltonian ${H}_{2}(t)$ in Eq.~(\ref{eq6}), it can be written as 
    \begin{equation}
    	\begin{aligned}
    		H_{I}=\sum_{n=-\infty }^{+\infty}h_{1,n}e^{i\omega_{1,n}t}+\sum_{m=-\infty }^{+\infty}h_{2,m}e^{i\omega_{2,m}t}+{\rm H.c.},
    	\end{aligned}
    \end{equation} 
    where the identification $h_{1,n}=\lambda J_{n}a^{\dagger}\sigma_{+}$, $h_{2,m}=\lambda J_{m}a\sigma_{+}$, $\omega_{1,n}=\Delta+(n+1)\omega_{f}$, and $\omega_{2,m}=\delta+m\omega_{f}$. Under the condition of the three-photon resonance: $\omega_{c} \approx \omega_{a}/3$, we have $\Delta\approx2\delta$ (i.e., $\omega_{1,-1}\approx2\omega_{2,0}$). When $\lvert\lambda J_{n} \rvert\ll \delta$,
    we can get the second-order effective Hamiltonian~\cite{PhysRevLett.85.2392,James2007}
    \begin{equation}
    	H_{{\rm eff}}^{(2)}=\frac{\lambda_{2}^{2}}{\delta}(a^{\dagger}a\sigma_z+\sigma_{+}\sigma_{-})+\frac{\lambda_{1}^{2}}{\Delta}(a^{\dagger}a\sigma_z-\sigma_{-}\sigma_{+})\\
    \end{equation}
    and the third-order effective Hamiltonian~\cite{Shao2017}
    \begin{equation}
    	H_{{\rm eff}}^{(3)}=-\frac{\lambda_{2}^{2}\lambda_{1}}{\delta^{2}}[a^{3}\sigma_{+}+(a^{\dagger})^{3}\sigma_{-}],
    \end{equation}
    where $\lambda_1 = \lambda J_{-1}$ and $\lambda_2 = \lambda J_{0}$. Therefore, the total effective Hamiltonian is given by
    \begin{eqnarray}
    	\nonumber
    	H_{{\rm eff}}&=&\omega_{c}a^{\dagger}a+\frac{\omega_{0}}{2}\sigma_z+H_{{\rm eff}}^{(2)}+H_{{\rm eff}}^{(3)}\\\nonumber
    	&=&\omega_{c}a^{\dagger}a+\frac{\omega_{0}}{2}\sigma_z+\frac{\lambda_{2}^{2}}{\delta}(a^{\dagger}a\sigma_z+\sigma_{+}\sigma_{-})\\\nonumber
    	&&+\frac{\lambda_{1}^{2}}{\Delta}(a^{\dagger}a\sigma_z-\sigma_{-}\sigma_{+})-\frac{\lambda_{2}^{2}\lambda_{1}}{\delta^{2}}[a^{3}\sigma_{+}+(a^{\dagger})^{3}\sigma_{-}],\\ \label{eq16}
    \end{eqnarray}
    which descrbes the transition between $\ket{e,0}$ and $\ket{g,3}$, with Rabi frequency $\Omega_{{\rm eff}}$,
    \begin{equation}
    	\Omega_{{\rm eff}}=\frac{\sqrt{6}\lambda_{2}^{2}\lambda_{1}}{\delta^{2}}.\label{17}
    \end{equation}
    To check the validity of the obtained result, we then write the matrix form of the effective Hamiltonian $H_{{\rm eff}}$ in the subspace formed by $\ket{e,0}$ and $\ket{g,3}$:
    \begin{equation}
    	H_{{\rm eff}}=\begin{pmatrix}
    		\frac{\omega_{0}}{2}+\frac{\lambda_{2}^2}{\delta} & -\frac{\sqrt{6}\lambda_{2}^2\lambda_{1}}{\delta^{2}} \\[2ex]
    		-\frac{\sqrt{6}\lambda_{2}^2\lambda_{1}}{\delta^{2}} &3\omega_{c}-\omega_{0}-\frac{3\lambda_{2}^2}{\delta}-\frac{4\lambda_{1}^2}{\Delta}
    	\end{pmatrix}.
    \end{equation}
    By equating the diagonal elements of this matrix, we can get a solution of the cavity field frequency $\omega_{c}=\omega_{c}^{'}$,
    \begin{equation}
    	\frac{\omega_{c}^{'}}{\omega_{0}}=\frac{1}{3}+2(J_{0}^{2}+\frac{1}{2}J_{-1}^{2})(\frac{\lambda}{\omega_{0}})^{2}+O(\frac{\lambda}{\omega_{0}})^{4}. \label{18}
    \end{equation}
    According to the effective Hamiltonian in Eq.~($\ref{eq16}$), $2\Omega_{{\rm eff}}$ can be understand as the energy splitting at the avoided crossing~\cite{16Ma2015}. We compare the energy splitting $\Delta E=2\Omega_{{\rm eff}}$ with the numerical simulation results obtained from the diagonalization of the Hamiltonian $H_{aR}$ when a resonant transition occurs between $\ket{e,0}$ and $\ket{g,3}$. Figure~$\ref{F7}$ shows the comparison between the analytical and numerical results for the energy splitting $\Delta E$ as a function of $\lambda/\omega_{0}$. It can be seen that the percentage difference is less than $3\%$ for $\lambda/\omega_{0}<0.05$. 
    
    \section{Output photon flux} \label{s5}
    In this section, we will use the master equation to study the output photon flux of the cavity. Cavity field damping and atomic decay caused by environmental noise inevitably have an impact on the evolution of the research object. We consider a system that the artificial atom and the cavity field are connected respectively to different baths. We assume that the temperature of both baths is zero. By using Born-Markov approximation, the master equation~\cite{PhysRevLett.109.193602,PhysRevLett.110.243601,PhysRevA.89.033827} for the reduced density matrix of the system $\rho(t)$ in the Schr\" {o}dinger picture can be written as
    \begin{equation}
    	\frac{d\rho(t)}{dt}=-i[H(t),\rho(t)]+\kappa D[X_{1}]\rho+\gamma D[X_{2}]\rho,     	
    \end{equation}
    where $H(t)$ is the Hamiltonian of the system given by Eq.$(\ref{eq1})$, the decay rates of the artificial atom and the bosonic mode are $\gamma$ and $\kappa$, respectively, and $X_{1}(t)$ and $X_{2}(t)$ are defined by
    \begin{equation}
    	\begin{aligned}
    		X_{1}&=\sum_{E_{n}>E_{m}}\bra{\psi_{m}}(a+a^{\dagger})\ket{\psi_{n}}\ket{\psi_{m}}\bra{\psi_{n}},\\
    		X_{2}&=\sum_{E_{n}>E_{m}}\bra{\psi_{m}}\sigma_{x}\ket{\psi_{n}}\ket{\psi_{m}}\bra{\psi_{n}},
    	\end{aligned}
    \end{equation}
    with $\ket{\psi_{n}}$ being the eigenvector of the Rabi Hamiltonian, i.e., $H_{R}\ket{\psi_{n}}=E_{n}\ket{\psi_{n}}$. The jumping operators are $X_{1}$ and $X_{2}$, respectively, which only contain transitions from high-energy eigenstates to low-energy eigenstates. The standard Lindblad superoperator $D$ is defined by
    \begin{equation}
    	D[O]\rho=\frac{1}{2}(2O\rho O^{\dagger}-O^{\dagger} O\rho-\rho O^{\dagger} O).
    \end{equation}
    The output photon flux rate~\cite{PhysRevLett.110.243601,PhysRevA.89.033827} from the cavity field is defined by
    \begin{equation}
    	\Phi_{{\rm out}}(t)=\kappa Tr[\rho(t)X_{1}^{\dagger}X_{1}].
    \end{equation}
    Note that the forms of $X_{1}$ and $X_{2}$ have taken the CR terms into account. If the Hamiltonian $H(t)$ is replaced by the Jaynes-Cummings (JC) Hamiltonian under the RWA, $X_{1}$ and $X_{2}$ are reduced to the forms $a$ and $a^{\dagger}$ respectively appearing in the standard Lindblad form master equation~\cite{book1}.
        \begin{figure}
    	\centering
    	\includegraphics[scale=0.43]{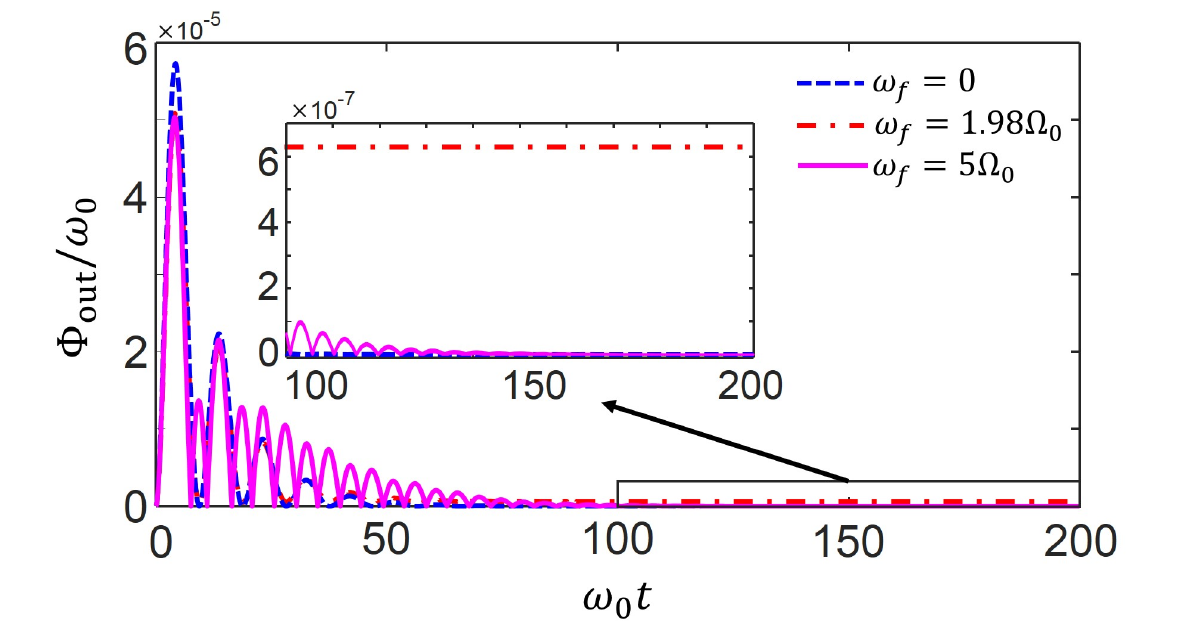}
    	\caption{Time dependence of the output photon flux rate $\Phi_{{\rm out}}(t)$ in various modulated cases. The inset in the panel shows the detail of the steady output flux rate. Parameters are $\lambda=0.01\omega_{0}$, $\kappa=\gamma=0.1\Omega_{0}$, $\Omega_{0}=100\omega_{0}$, and $x=0.5$.}
    	\label{F9}
    \end{figure} 
    We assume that the system is initially prepared in the state $\ket{e,0}$, and obtain the reduced density operator $\rho(t)$ of the system by numerically solving the master equation. In Fig.~$\ref{F9}$, we plot the time dependence of the output bosonic excitation flux rate $\Phi_{{\rm out}}(t)$. For the driving frequency $\omega_{f}=0$, there is no steady output flux. After applying the drive with frequency $\omega_{f}=1.98\Omega_{0}$ and amplitude $A=0.99\Omega_{0}$, we can see that the value of $\Phi_{{\rm out}}(t)$ approaches a stationary value when $t\gg100/\omega_{0}$. When the modulation frequency is increased to $5\Omega_{0}$, the steady output photon flux rate disappears. By analyzing the form of the Hamiltonian describing the system, we can explain why such a stable output bosonic excitation flux occurs only when $\omega_{f}\sim 2\Omega_{0}$. When there is no external driving, the system can be well described by the JC model in the near-resonance regime (i.e. the transition frequency of the atom and the cavity frequency are almost equal). If the system is initially prepared in the state $\ket{e,0}$, the dissipation induced by the environment (the zero-temperature baths) will lead the system to the lowest eigenstates $\ket{g,0}$, and the system will always stay in this state without emitting stable output photon flux. As the appropriate modulation is applied, the system is described by the quantum Rabi model, and the lowest energy eigenstate is no longer the state $\ket{g,0}$. Instead, the state $\ket{g,0}$ is a superposition of many eigenstates of the Rabi model. The influence of cavity field damping and atomic decay will induce transitions from high-energy states to low-energy states resulting in a stable output photon flux rate. 
    
    We show that when the modulation frequency is $\omega_{f}=1.98\Omega_{0}$, the system generates a three-photon Rabi oscillation. At this frequency, the dynamics of the system is well described by an effective anisotropic Rabi model, so that the system continues to radiate a stream of photons outward when dissipation is considered. However, when the modulation frequency is equal to $5\Omega_{0} $ or higher, the CR terms in Eq.~\ref{eq9} can be neglected, at which point the dynamics of the system is described by the JC model, so no steady output flux is produced.
    
    \section{Discussion on the influence of Energy Level Anharmonicity} \label{s4}
\begin{figure*}
  	\centering
  	\subfigure{
  		\label{F8(a)}
  		\includegraphics[scale=0.42]{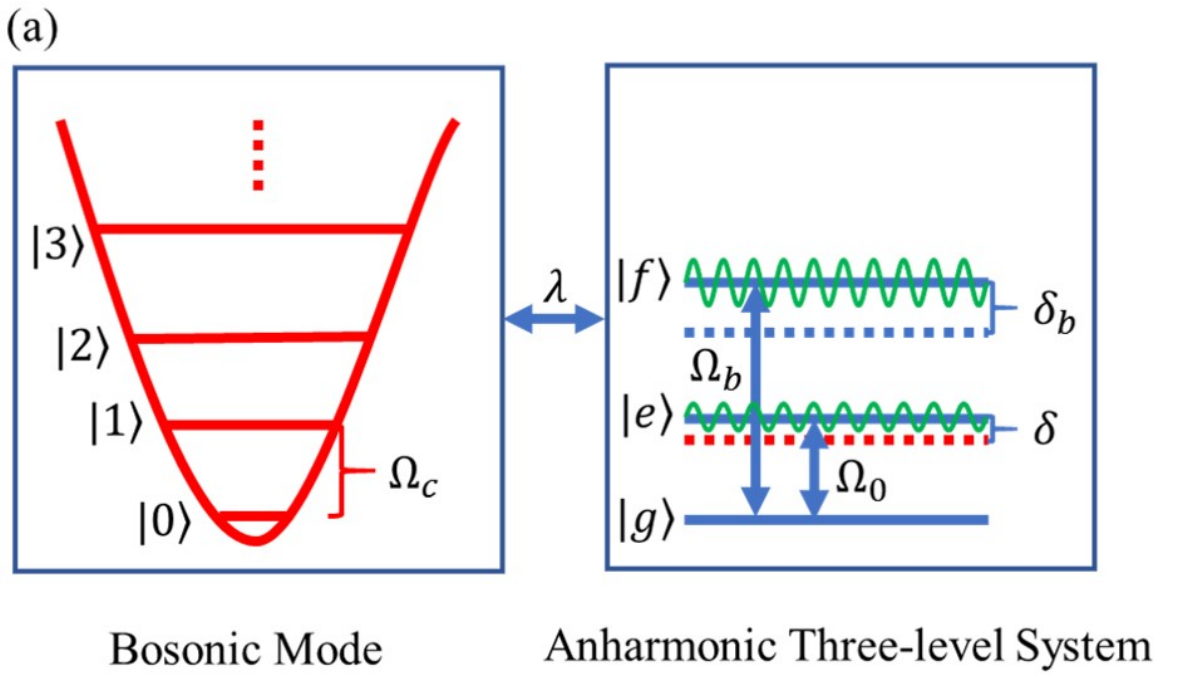}
  	}
  	\subfigure{
  		\label{F8(b)}
  		\includegraphics[scale=0.42]{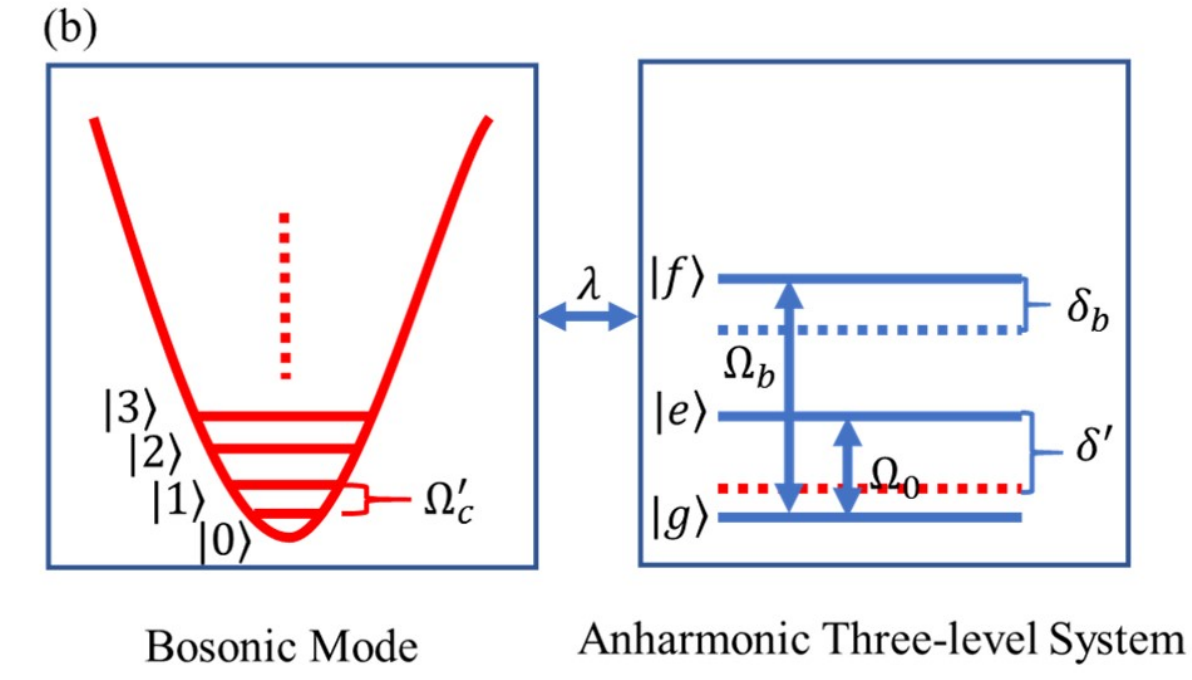}
  	}
  	
  	\caption{Sketch for the coupling between a bosonic mode and a three-level system. The transition frequency between $\ket{g}$ and $\ket{e}$ as well as $\ket{e}$ and $\ket{f}$ are $\Omega_{0}$ and $\Omega_{b}$, respectively. The parameter $\delta_{b}$ denotes the energy level anharmonicity and the parameter $\delta$ denotes the detuning of the bosonic mode frequency and the transition frequency between $\ket{g}$ and $\ket{e}$. In the theoretical model (a), the frequency of the bosonic mode is $\Omega_{c}$. A sine modulation is applied to the state $\ket{e}$ and $\ket{f}$. In the theoretical model (b), the bosonic mode frequency $\Omega_{c}^{'} \approx \Omega_{0}/3$ and the three-level system is not driven by an external field.}
  	\label{F8}
  \end{figure*}

    For simplicity, we consider the lowest three energy levels of the artificial atom that is biased by a sinusoidal flux (see Fig.~\ref{F8(a)}). The system dynamics is described by the Hamiltonian ($\hbar=1$)
    \begin{equation}
    	\begin{aligned}
    		H_{3}(t)=[&\Omega_{0}+A\cos(\omega_{f}t)]\ket{e}\bra{e}\\
    		+&[\Omega_{b}+2A\cos(\omega_{f}t)]\ket{f}\bra{f}\\
    		+&\Omega_{c}a^{\dagger}a+\lambda(a^{\dagger}+a)(\ket{e}\bra{g}+\ket{f}\bra{e}+\rm H.c.). \label{eq22}
    	\end{aligned}
    \end{equation}
    In the rotating frame defined by
    \begin{equation}
    	\begin{aligned}
    		U_{2}(t)={\rm exp}& \Big[-i(\Omega_{c}a^{\dagger}a+\Omega_{0}\ket{e}\bra{e}+\Omega_{b}\ket{f}\bra{f})t\\
    		&-ix\cos(\omega_{f}t)\left(\ket{e}\bra{e}+2\ket{f}\bra{f}\right)\Big] ,
    	\end{aligned}
    \end{equation}
    the Hamiltonian in Eq.~(\ref{eq22}) is written as 
    \begin{equation}
    	\begin{aligned}
    		H_{3}^{'}(t)=
    		\lambda& \Big[a^{\dagger}\ket{e}\bra{g}e^{i(\Omega_{0}+\Omega_{c})t+ix\sin(\omega_{f}t)}\\
    		+&a\ket{e}\bra{g}e^{i(\Omega_{0}-\Omega_{c})t+ix\sin(\omega_{f}t)}\\
    		+&a^{\dagger}\ket{f}\bra{e}e^{i(\Omega_{b}-\Omega_{0}+\Omega_{c})t+ix\sin(\omega_{f}t)}\\
    		+&a\ket{f}\bra{e}e^{i(\Omega_{b}-\Omega_{0}-\Omega_{c})t+ix\sin(\omega_{f}t)}+\rm H.c.\Big],
    	\end{aligned}
    \end{equation}
    where $x=A/\omega_{f}$. By setting $\delta_{b}=\Omega_{b}-2\Omega_{0}$ as the anharmonicity of atomic energy levels, and using Eq.~(\ref{eq5}), resulting in
    \begin{equation}
    	\begin{aligned}
    		H_{3}^{''}(t)=\lambda& \left(J_{-1}a^{\dagger}\ket{e}\bra{g}e^{i\Delta t}+ J_{0}a\ket{e}\bra{g}e^{i\delta t}\right)\\    		
    		+&\sum_{n\neq-1} \lambda J_{n} a^{\dagger}\ket{e}\bra{g}e^{i[\Delta+(1+n)\omega_{f}]t}\\
    		+&\sum_{m\neq0} \lambda J_{m} a\ket{e}\bra{g}e^{i(\Delta+m\omega_{f})t}\\
    		+&\sum_{l} \lambda J_{l}
    		a^{\dagger}\ket{f}\bra{e}e^{i[\delta_{b}+\Delta+(1+l)\omega_{f}]t}\\
    		+&\sum_{k} \lambda J_{k}
    		\ket{f}\bra{e}e^{i[\delta_{b}+\delta+k\omega_{f}]t}
    		+{\rm H.c.} \label{eq25}
    	\end{aligned}
    \end{equation}
    Under the conditions
    \begin{equation}
    	\lambda \lvert J_{n}(x) \rvert \ll \delta \ll \{\omega_{f},\delta_{b}\}, 
    \end{equation}
    the Hamiltonian $H_{3}^{''}(t)$ in Eq.~(\ref{eq25}) can be simplified as
    \begin{equation}
    	\begin{aligned}
    		\widetilde{H}_{3}(t)\approx \lambda J_{1}a^{\dagger}\ket{e}\bra{g}e^{i\Delta t}
    		+\lambda J_{0}a\ket{e}\bra{g}e^{i\delta t}+\rm H.c.,
    		\label{eq24}
    	\end{aligned}
    \end{equation}
    which is equivalent to the Hamiltonian in Eq.~(\ref{eq9}). Note that, in the protocol, the detuning $\delta$ can be very small, such that the anharmonicity $\delta_{b}$ can be much greater than $\delta$. It shows that even considering the influence of high energy levels, the protocol can also generate the three-photon resonance.
    
    Without the frequency modulation (see Fig.~\ref{F8(b)}), in the rotating frame, the Hamiltonian of the QRM is given by
    \begin{equation}
    	\begin{aligned}
    		H_{4}(t)=
    		\lambda& \Big[a^{\dagger}\ket{e}\bra{g}e^{i\Delta t}+a\ket{e}\bra{g}e^{i\delta^{'} t}\\
    		+&a^{\dagger}\ket{f}\bra{e}e^{i(\Delta+\delta_{b})t}+a\ket{f}\bra{e}e^{i(\delta^{'}+\delta_{b})t}+\rm H.c.\Big].
    		\label{eq25}
    	\end{aligned}
    \end{equation}
    To avoid the excitation of the second-excited state $\ket{f}$, according to Eq.~(\ref{eq25}), the condition $\delta_{b}\gg \delta^{'}$ must be well satisfied. Due to the large-detuning regime, the anharmonicity $\delta_{b}$ and the detuning $\delta^{'}$ are on the same order of magnitude, which makes the second row in Eq.~(\ref{eq25}) unable to be omitted by performing RWA. Therefore, the existence of the third excited state $\ket{f}$ causes the three-photon resonance not to occur.
    \section{circuit implementation and the experimental parameters} \label{s6}
    \begin{figure}{H}
    	\centering
    	\includegraphics[scale=0.43]{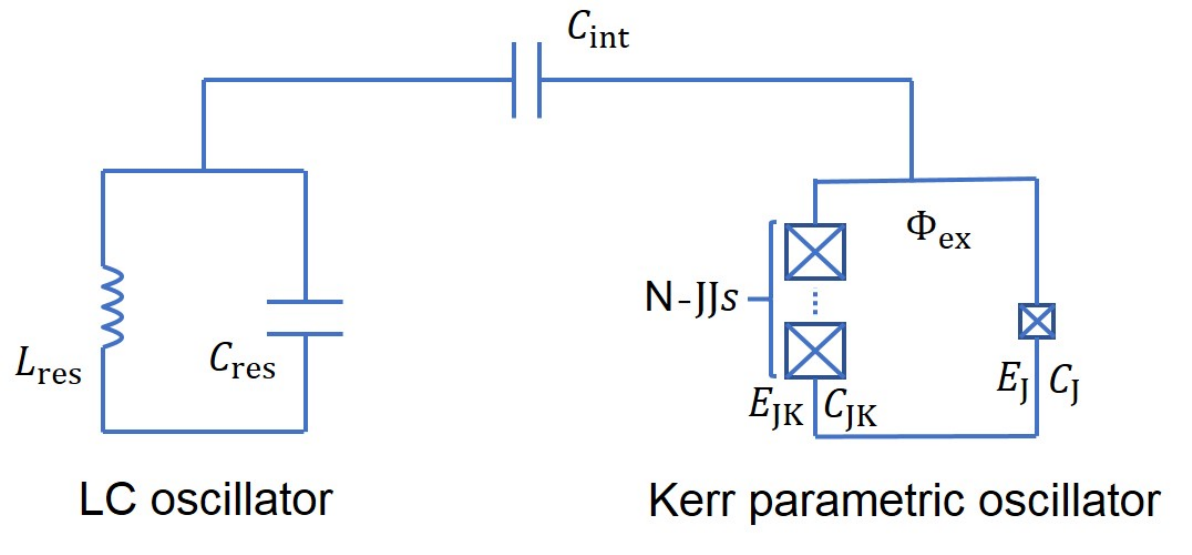}
    	\caption{Circuit diagram of an $LC$ oscillator and a Kerr parametric oscillator (artificial atom). JJ stands for “Josephson junction”. Two large junctions with
    		vertical dots represent a junction array. The junction capacitances is not shown for simplicity.}
    	\label{F10}
    \end{figure}
    The protocol has the potential to be implemented in the circuit in Fig.~\ref{F10}. After quantisation (see Appendix~\ref{app:A}), the $LC$ oscillator can be represented as a bosonic field and the Kerr parametric oscillator (KPO) as an artificial atom with modulated energy levels separation. The applied flux bias is $\Phi_{{\rm ex}}=A\cos(\omega_{f}t)$ with $A=0.5\omega_{f}$ and $\omega_{f}=9.93$ GHz. As reported in Ref.~\cite{li_motional_2013}, the modulation frequency $\omega_{f}/2\pi$ can range from 0 to 500 MHz in a circuit-QED, leading the modulation frequency close to the same order of magnitude as the atomic transition frequency. Therefore, it is possible that the modulation frequency may approach twice the artificial atom transition frequency. In the protocol, the frequency of the cavity field is $\Omega_{c}=5$ GHz, the atomic transition frequency is $\Omega_{0}=5.066$ GHz, and the coupling strength $\lambda=0.198$ MHz. The artificial atom decaying rate $\gamma$ and the cavity field decaying rate $\kappa$ are both 1.98 MHz. The above parameters are all experimentally reasonable~\cite{8WOS:000366730900005,PhysRevLett.96.127006,RevModPhys.91.025005,10WOS:001112616000002,li_motional_2013,PhysRevLett.131.113601}. By simulating the parameters in the protocol, we can observe the steady photon flux rate $\Phi_{{\rm out}}=0.06$ kHz that can be detected in experiments.
      
    \section{CONCLUSIONS} \label{s7}
    We propose a protocol to generate three-photon Rabi oscillations in the small-detuning regime. Specifically, by introducing a modulation to the artificial atom in the Rabi model, we can obtain an effective anisotropic Rabi model with tunable parameters. Through numerical simulations, clear evidence is shown that the three-photon resonance occurs when the modulation frequency is close to twice the transition frequency of the artificial atom. We derive the three-photon coupling effective Hamiltonian between $\ket{e,0}$ and $\ket{g,3}$, which determines the resonance position and the energy splitting in the coupling regime $\lambda<0.05\omega_{0}$. The discussion of energy level anharmonicity indicates that the protocol can work well with small anharmonicity. Additionally, we calculate the output photon flux of the entire system. It is worth noting that there exists a nonzero stationary value of the output bosonic excitation flux rate as we apply a modulation to the artifical atom, indicating the occurrence of the three-photon resonance. The protocol will provide a way for investigating the excitation-number-nonconserving processes and have a more profound impact in the near future.
     
    \section*{Acknowledgements}
    Y.X. was supported by the National Natural Science Foundation of China under Grants No. 11575045 and No. 11874114, the Natural Science Funds for Distinguished Young Scholar of Fujian Province under Grant No. 2020J06011, and a project from Fuzhou University under Grant No.  JG2020001-2.  Y.-H.C. was supported by the National Natural Science Foundation of China under Grant No. 12304390.
	
	\appendix
	\section{Derivation of the Hamiltonian} \label{app:A}
	In this appednix, we derive the Hamiltonian of the circuit in Fig.~\ref{F10}. The Lagrangian of the circuit can be written as 
	\begin{equation}
		\mathcal{L}=\mathcal{T}-\mathcal{U},
	\end{equation}
	where
	\begin{equation}
		\begin{aligned}
			\mathcal{T}&=\left(\frac{\Phi_{0}}{2\pi}\right)^{2}\left[(C_{\rm J}+\frac{C_{\rm JK}}{2N})\dot{\phi}^2+\frac{C_{\rm res}}{2}\dot{\theta}^2+\frac{C_{\rm i}}{2}(\dot{\phi}+\dot{\theta})^2\right], \\ 
			\mathcal{U}&=-E_{\rm J}\cos(\phi+\frac{\varphi_{\rm ex}}{2})-E_{\rm J}\cos(\phi-\frac{\varphi_{\rm ex}}{2})\\
			&\ \ \ \ -NE_{\rm JK}\cos(\frac{\phi}{N})+\left(\frac{\Phi_{0}}{2\pi}\right)^{2}\frac{1}{2L_{\rm res}}\theta^2.  		
		\end{aligned}
	\end{equation}
	Here, $\Phi_{0}$ is the magnetic flux quantum and $\varphi_{\rm ex}\equiv 2\pi\Phi_{{\rm ex}}/\Phi_{0}$; $E_{\rm J}$ is the Josephson energy and $C_{\rm J}$ is the capacitance of each junction in the DC SQUID part of the KPO; the energy and capacitance for each junction in the junction array are $E_{\rm JK}$ and $C_{\rm JK}$; $L_{\rm res}$ is the inductance and $C_{\rm res}$ is the capacitance of the $LC$ resonator; and $C_{\rm i}$ is the coupling capacitance between the KPO and the $LC$ resonator. By defining the conjugate number operators as 
	\begin{equation}
		\begin{aligned}
			\hbar \hat{N}_{\phi}&=\frac{\partial \mathcal{L}}{\partial \dot{\phi}}=C_{\rm KPO}\dot{\phi}-C_{\rm int}\dot{\theta},\\
			\hbar \hat{N}_{\theta}&=\frac{\partial \mathcal{L}}{\partial \dot{\theta}}=-C_{\rm int}\dot{\phi}+C_{\rm an}\dot{\theta},
		\end{aligned}
	\end{equation}
	where
	$C_{\rm KPO}\equiv (\Phi_{0}/2\pi)^2(2C_{\rm J}+C_{\rm JK}/N+C_{\rm int})$, $C_{\rm an}\equiv (\Phi_{0}/2\pi)^2(C_{\rm res}+C_{\rm int})$, and $C_{\rm int}\equiv (\Phi_{0}/2\pi)^2C_{\rm i}$, the resulting Hamiltonian is 
	\begin{equation}
		\begin{aligned}
			\hat{\mathcal{H}}=
			4&E_{\rm C}^{\phi}\hat{N}_{\phi}^2+4E_{\rm C}^{\theta}\hat{N}_{\theta}^2+8E_{\rm C}^{\rm int}\hat{N}_{\phi}\hat{N}_{\theta}+\frac{E_{\rm L}}{2}\hat{\theta}^2\\
			&-2E_{\rm J}\cos\left(\frac{\varphi_{\rm ex}}{2}\right)\cos(\hat{\phi})
			-NE_{\rm JK}\cos\left(\frac{\hat{\phi}}{N}\right),
		\end{aligned}
	\end{equation}
	where 
	\begin{eqnarray}
		\nonumber
		&E_{\rm C}^{\phi}\equiv\frac{\hbar^2C_{\rm an}}{8(C_{\rm KPO}C_{\rm an}-C_{\rm int}^2)}, E_{\rm C}^{\theta}\equiv\frac{\hbar^2C_{\rm KPO}}{8(C_{\rm KPO}C_{\rm an}-C_{\rm int}^2)},&\\ \nonumber
		&E_{\rm C}^{\rm int}\equiv\frac{\hbar^2C_{\rm int}}{8(C_{\rm KPO}C_{\rm an}-C_{\rm int}^2)}, E_{\rm C}^{\rm int}\equiv\frac{\hbar^2C_{\rm int}}{8(C_{\rm KPO}C_{\rm an}-C_{\rm int}^2)},&\\ \nonumber
		&E_{\rm L}\equiv (\Phi_{0}/2\pi)^2\frac{1}{L_{\rm res}}.& \nonumber
	\end{eqnarray}
	We move to the occupation-number representation by defining
	\begin{eqnarray}
		\nonumber
		&\hat{N}_{\phi}=iN_{0}^{\phi}(\hat{a}^{\dagger}-\hat{a}), \hat{\phi}=\phi_{0}(\hat{a}^{\dagger}+\hat{a}),&\\\nonumber
		&\hat{N}_{\theta}=iN_{0}^{\theta}(\hat{b}^{\dagger}-\hat{b}),
		\hat{\theta}=\theta_{0}(\hat{b}^{\dagger}+\hat{b}),&\\
		\label{A5}
	\end{eqnarray}
	where $N_{0}^{\phi}=\sqrt[4]{E_{\rm JK}/32NE_{\rm C}^{\phi}}$, $\phi_{0}=\sqrt[4]{2NE_{\rm C}^{\phi}/E_{\rm JK}}$, $N_{0}^{\theta}=\sqrt[4]{E_{\rm L}/32NE_{\rm C}^{\theta}}$, and $\theta_{0}=\sqrt[4]{2E_{\rm C}^{\theta}/E_{\rm L}}$, we have 
	\begin{equation}
		\begin{aligned}
			\hat{\mathcal{H}}=\hbar&\omega_{\rm KPO}\hat{a}^{\dagger}a+\hbar\omega_{\rm LC}\hat{b}^{\dagger}b+\hbar\lambda(\hat{a}^{\dagger}\hat{b}+\hat{a}\hat{b}^{\dagger}-\hat{a}\hat{b}-\hat{a}^{\dagger}\hat{b}^{\dagger})\\
			-&NE_{\rm JK}\cos\left(\frac{\hat{\phi}}{N}\right)-\frac{E_{\rm JK}}{2N}\hat{\phi}^{2}-2E_{\rm J}\cos\left(\frac{\varphi_{\rm ex}}{2}\right)\cos(\hat{\phi}),
		\end{aligned}
	\end{equation}
	where $\hbar\omega_{\rm KPO}\equiv\sqrt{8E_{\rm C}^{\rm \phi}E_{\rm JK}/N}$, $\hbar\omega_{\rm LC}\equiv\sqrt{8E_{\rm C}^{\rm \theta}E_{\rm L}}$, and $\hbar\lambda\equiv E_{\rm C}^{\rm int}\sqrt[4]{4E_{\rm JK}E_{\rm L}/(NE_{\rm C}^{\rm \phi}E_{\rm C}^{\rm \theta})}$. 
	
	Using the Taylor series $\cos\left(\hat{\phi}\right)=1-\frac{\hat{\phi}^2}{2}+\frac{\hat{\phi}^4}{24}...$, we have 
	\begin{eqnarray}
		\nonumber
		\hat{\mathcal{H}}_{\rm JJ}&=&
		-NE_{\rm JK}\left[1+\frac{1}{24}\left(\frac{\hat{\phi}}{N}\right)^{4}-...\right]\\ \nonumber
		&&-2E_{\rm J}\cos\left(\frac{\varphi_{\rm ex}}{2}\right)\left(1-\frac{\hat{\phi}^2}{2}+\frac{\hat{\phi}^4}{24}-...\right),\\
	\end{eqnarray}
	which can lead to energy level anharmonicity and energy level modulation. Utilizing the Eq.~(\ref{A5}), we can obtain the Hamiltonian for the circuit as follows,
	\begin{eqnarray}
		\nonumber
		\hat{\mathcal{H}}&\approx&[\hbar\omega_{\rm KPO}-\frac{E_{\rm C}^{\rm \phi}}{N^{2}}+\frac{\hbar\omega_{\rm KPO}NE_{\rm J}}{E_{\rm JK}}\cos\left(\frac{\varphi_{\rm ex}}{2}\right)]\hat{a}^{\dagger}\hat{a}\\ \nonumber
		&&-\frac{E_{\rm C}^{\rm \phi}}{2N^{2}}\hat{a}^{\dagger}\hat{a}^{\dagger}\hat{a}\hat{a}+\hbar\omega_{\rm LC}\hat{b}^{\dagger}b+\hbar\lambda(\hat{a}^{\dagger}\hat{b}+\hat{a}\hat{b}^{\dagger}-\hat{a}\hat{b}-\hat{a}^{\dagger}\hat{b}^{\dagger})\\ \nonumber
		&=&\hbar\left[\Omega_{0}+A\cos(\omega_{f}t)\right]\hat{a}^{\dagger}\hat{a}+\hbar\Omega_{c}\hat{b}^{\dagger}\hat{b}-\hbar\delta_{b}\hat{a}^{\dagger}\hat{a}^{\dagger}\hat{a}\hat{a}\\ \nonumber
		&&+\hbar\lambda(\hat{a}^{\dagger}\hat{b}+\hat{a}\hat{b}^{\dagger}-\hat{a}\hat{b}-\hat{a}^{\dagger}\hat{b}^{\dagger}),\\ 
	\end{eqnarray}
    where $\hbar\Omega_{0}=\hbar\omega_{\rm KPO}-\frac{E_{\rm C}^{\phi}}{N^{2}}$, $\hbar A=\frac{\hbar \omega_{\rm KPO}NE_{\rm J}}{E_{\rm JK}}$, $\omega_{f}t=\frac{\varphi_{\rm ex}}{2}$, $\hbar \Omega_{c}=\hbar \omega_{\rm LC}$, and $\hbar \delta_{b}=\frac{E_{\rm C}^{\phi}}{N^{2}}$. By controlling the value of the inharmonicity $\delta_{b}$, we can consider the KPO as an artificial atom with only the lowest three energy levels. Therefore, we obtain the Hamiltonian in Eq.~(\ref{eq18}).

	\bibliography{reference}

\begin{thebibliography}{59}%
\makeatletter
\providecommand \@ifxundefined [1]{%
 \@ifx{#1\undefined}
}%
\providecommand \@ifnum [1]{%
 \ifnum #1\expandafter \@firstoftwo
 \else \expandafter \@secondoftwo
 \fi
}%
\providecommand \@ifx [1]{%
 \ifx #1\expandafter \@firstoftwo
 \else \expandafter \@secondoftwo
 \fi
}%
\providecommand \natexlab [1]{#1}%
\providecommand \enquote  [1]{``#1''}%
\providecommand \bibnamefont  [1]{#1}%
\providecommand \bibfnamefont [1]{#1}%
\providecommand \citenamefont [1]{#1}%
\providecommand \href@noop [0]{\@secondoftwo}%
\providecommand \href [0]{\begingroup \@sanitize@url \@href}%
\providecommand \@href[1]{\@@startlink{#1}\@@href}%
\providecommand \@@href[1]{\endgroup#1\@@endlink}%
\providecommand \@sanitize@url [0]{\catcode `\\12\catcode `\$12\catcode
  `\&12\catcode `\#12\catcode `\^12\catcode `\_12\catcode `\%12\relax}%
\providecommand \@@startlink[1]{}%
\providecommand \@@endlink[0]{}%
\providecommand \url  [0]{\begingroup\@sanitize@url \@url }%
\providecommand \@url [1]{\endgroup\@href {#1}{\urlprefix }}%
\providecommand \urlprefix  [0]{URL }%
\providecommand \Eprint [0]{\href }%
\providecommand \doibase [0]{https://doi.org/}%
\providecommand \selectlanguage [0]{\@gobble}%
\providecommand \bibinfo  [0]{\@secondoftwo}%
\providecommand \bibfield  [0]{\@secondoftwo}%
\providecommand \translation [1]{[#1]}%
\providecommand \BibitemOpen [0]{}%
\providecommand \bibitemStop [0]{}%
\providecommand \bibitemNoStop [0]{.\EOS\space}%
\providecommand \EOS [0]{\spacefactor3000\relax}%
\providecommand \BibitemShut  [1]{\csname bibitem#1\endcsname}%
\let\auto@bib@innerbib\@empty
\bibitem [{\citenamefont {Rabi}(1936)}]{PhysRev.49.324}%
  \BibitemOpen
  \bibfield  {author} {\bibinfo {author} {\bibfnamefont {I.~I.}\ \bibnamefont
  {Rabi}},\ }\bibfield  {title} {\bibinfo {title} {On the process of space
  quantization},\ }\href {https://doi.org/10.1103/PhysRev.49.324} {\bibfield
  {journal} {\bibinfo  {journal} {Phys. Rev.}\ }\textbf {\bibinfo {volume}
  {49}},\ \bibinfo {pages} {324} (\bibinfo {year} {1936})}\BibitemShut
  {NoStop}%
\bibitem [{\citenamefont {Rabi}(1937)}]{4PhysRev.51.652}%
  \BibitemOpen
  \bibfield  {author} {\bibinfo {author} {\bibfnamefont {I.~I.}\ \bibnamefont
  {Rabi}},\ }\bibfield  {title} {\bibinfo {title} {Space quantization in a
  gyrating magnetic field},\ }\href {https://doi.org/10.1103/PhysRev.51.652}
  {\bibfield  {journal} {\bibinfo  {journal} {Phys. Rev.}\ }\textbf {\bibinfo
  {volume} {51}},\ \bibinfo {pages} {652} (\bibinfo {year} {1937})}\BibitemShut
  {NoStop}%
\bibitem [{\citenamefont {Brune}\ \emph {et~al.}(1996)\citenamefont {Brune},
  \citenamefont {Schmidt-Kaler}, \citenamefont {Maali}, \citenamefont {Dreyer},
  \citenamefont {Hagley}, \citenamefont {Raimond},\ and\ \citenamefont
  {Haroche}}]{1PRL76.1800}%
  \BibitemOpen
  \bibfield  {author} {\bibinfo {author} {\bibfnamefont {M.}~\bibnamefont
  {Brune}}, \bibinfo {author} {\bibfnamefont {F.}~\bibnamefont
  {Schmidt-Kaler}}, \bibinfo {author} {\bibfnamefont {A.}~\bibnamefont
  {Maali}}, \bibinfo {author} {\bibfnamefont {J.}~\bibnamefont {Dreyer}},
  \bibinfo {author} {\bibfnamefont {E.}~\bibnamefont {Hagley}}, \bibinfo
  {author} {\bibfnamefont {J.~M.}\ \bibnamefont {Raimond}},\ and\ \bibinfo
  {author} {\bibfnamefont {S.}~\bibnamefont {Haroche}},\ }\bibfield  {title}
  {\bibinfo {title} {Quantum {Rabi} oscillation: A direct test of field
  quantization in a cavity},\ }\href
  {https://doi.org/10.1103/PhysRevLett.76.1800} {\bibfield  {journal} {\bibinfo
   {journal} {Phys. Rev. Lett.}\ }\textbf {\bibinfo {volume} {76}},\ \bibinfo
  {pages} {1800} (\bibinfo {year} {1996})}\BibitemShut {NoStop}%
\bibitem [{\citenamefont {Braak}(2011)}]{PhysRevLett.107.100401}%
  \BibitemOpen
  \bibfield  {author} {\bibinfo {author} {\bibfnamefont {D.}~\bibnamefont
  {Braak}},\ }\bibfield  {title} {\bibinfo {title} {Integrability of the {Rabi}
  model},\ }\href {https://doi.org/10.1103/PhysRevLett.107.100401} {\bibfield
  {journal} {\bibinfo  {journal} {Phys. Rev. Lett.}\ }\textbf {\bibinfo
  {volume} {107}},\ \bibinfo {pages} {100401} (\bibinfo {year}
  {2011})}\BibitemShut {NoStop}%
\bibitem [{\citenamefont {Ashhab}\ \emph {et~al.}(2006)\citenamefont {Ashhab},
  \citenamefont {Johansson},\ and\ \citenamefont {Nori}}]{ashhab_rabi_2006}%
  \BibitemOpen
  \bibfield  {author} {\bibinfo {author} {\bibfnamefont {S.}~\bibnamefont
  {Ashhab}}, \bibinfo {author} {\bibfnamefont {J.~R.}\ \bibnamefont
  {Johansson}},\ and\ \bibinfo {author} {\bibfnamefont {F.}~\bibnamefont
  {Nori}},\ }\bibfield  {title} {\bibinfo {title} {{Rabi} oscillations in a
  qubit coupled to a quantum two-level system},\ }\href
  {https://doi.org/10.1088/1367-2630/8/6/103} {\bibfield  {journal} {\bibinfo
  {journal} {New J. Phys.}\ }\textbf {\bibinfo {volume} {8}},\ \bibinfo {pages}
  {103} (\bibinfo {year} {2006})}\BibitemShut {NoStop}%
\bibitem [{\citenamefont {Auffeves}\ \emph {et~al.}(2013)\citenamefont
  {Auffeves} \emph {et~al.}}]{kwek2013strong}%
  \BibitemOpen
  \bibfield  {author} {\bibinfo {author} {\bibfnamefont {A.}~\bibnamefont
  {Auffeves}} \emph {et~al.},\ }\href@noop {} {\emph {\bibinfo {title} {Strong
  light-matter coupling: from atoms to solid-state systems}}}\ (\bibinfo
  {publisher} {World Scientific},\ \bibinfo {year} {2013})\BibitemShut
  {NoStop}%
\bibitem [{\citenamefont {Ning}\ \emph {et~al.}(2024)\citenamefont {Ning},
  \citenamefont {Zheng}, \citenamefont {Lü}, \citenamefont {Wu}, \citenamefont
  {Yang},\ and\ \citenamefont {Zheng}}]{2WOS:001142919400001}%
  \BibitemOpen
  \bibfield  {author} {\bibinfo {author} {\bibfnamefont {W.}~\bibnamefont
  {Ning}}, \bibinfo {author} {\bibfnamefont {R.-H.}\ \bibnamefont {Zheng}},
  \bibinfo {author} {\bibfnamefont {J.-H.}\ \bibnamefont {Lü}}, \bibinfo
  {author} {\bibfnamefont {F.}~\bibnamefont {Wu}}, \bibinfo {author}
  {\bibfnamefont {Z.-B.}\ \bibnamefont {Yang}},\ and\ \bibinfo {author}
  {\bibfnamefont {S.-B.}\ \bibnamefont {Zheng}},\ }\bibfield  {title} {\bibinfo
  {title} {Experimental observation of spontaneous symmetry breaking in a
  quantum phase transition},\ }\href
  {https://doi.org/10.1007/s11433-023-2259-1} {\bibfield  {journal} {\bibinfo
  {journal} {Sci. China Phys. Mech. Astron.}\ }\textbf {\bibinfo {volume}
  {67}},\ \bibinfo {pages} {220312} (\bibinfo {year} {2024})}\BibitemShut
  {NoStop}%
\bibitem [{\citenamefont {Chen}\ \emph
  {et~al.}(2024{\natexlab{a}})\citenamefont {Chen}, \citenamefont {Qiu},
  \citenamefont {Miranowicz}, \citenamefont {Lambert}, \citenamefont {Qin},
  \citenamefont {Stassi}, \citenamefont {Xia}, \citenamefont {Zheng},\ and\
  \citenamefont {Nori}}]{3WOS:001137118100005}%
  \BibitemOpen
  \bibfield  {author} {\bibinfo {author} {\bibfnamefont {Y.-H.}\ \bibnamefont
  {Chen}}, \bibinfo {author} {\bibfnamefont {Y.}~\bibnamefont {Qiu}}, \bibinfo
  {author} {\bibfnamefont {A.}~\bibnamefont {Miranowicz}}, \bibinfo {author}
  {\bibfnamefont {N.}~\bibnamefont {Lambert}}, \bibinfo {author} {\bibfnamefont
  {W.}~\bibnamefont {Qin}}, \bibinfo {author} {\bibfnamefont {R.}~\bibnamefont
  {Stassi}}, \bibinfo {author} {\bibfnamefont {Y.}~\bibnamefont {Xia}},
  \bibinfo {author} {\bibfnamefont {S.-B.}\ \bibnamefont {Zheng}},\ and\
  \bibinfo {author} {\bibfnamefont {F.}~\bibnamefont {Nori}},\ }\bibfield
  {title} {\bibinfo {title} {Sudden change of the photon output field marks
  phase transitions in the quantum {Rabi} model},\ }\href
  {https://doi.org/10.1038/s42005-023-01457-w} {\bibfield  {journal} {\bibinfo
  {journal} {Commun. Phys.}\ }\textbf {\bibinfo {volume} {7}},\ \bibinfo
  {pages} {5} (\bibinfo {year} {2024}{\natexlab{a}})}\BibitemShut {NoStop}%
\bibitem [{\citenamefont {Duan}(2023)}]{PhysRevB.108.174306}%
  \BibitemOpen
  \bibfield  {author} {\bibinfo {author} {\bibfnamefont {L.-W.}\ \bibnamefont
  {Duan}},\ }\bibfield  {title} {\bibinfo {title} {Periodic jumps in binary
  lattices with a static force},\ }\href
  {https://doi.org/10.1103/PhysRevB.108.174306} {\bibfield  {journal} {\bibinfo
   {journal} {Phys. Rev. B}\ }\textbf {\bibinfo {volume} {108}},\ \bibinfo
  {pages} {174306} (\bibinfo {year} {2023})}\BibitemShut {NoStop}%
\bibitem [{\citenamefont {Mei}\ \emph {et~al.}(2022)\citenamefont {Mei},
  \citenamefont {Li}, \citenamefont {Wu}, \citenamefont {Cai}, \citenamefont
  {Wang}, \citenamefont {Yao}, \citenamefont {Zhou},\ and\ \citenamefont
  {Duan}}]{PhysRevLett.128.160504}%
  \BibitemOpen
  \bibfield  {author} {\bibinfo {author} {\bibfnamefont {Q.-X.}\ \bibnamefont
  {Mei}}, \bibinfo {author} {\bibfnamefont {B.-W.}\ \bibnamefont {Li}},
  \bibinfo {author} {\bibfnamefont {Y.-K.}\ \bibnamefont {Wu}}, \bibinfo
  {author} {\bibfnamefont {M.-L.}\ \bibnamefont {Cai}}, \bibinfo {author}
  {\bibfnamefont {Y.}~\bibnamefont {Wang}}, \bibinfo {author} {\bibfnamefont
  {L.}~\bibnamefont {Yao}}, \bibinfo {author} {\bibfnamefont {Z.-C.}\
  \bibnamefont {Zhou}},\ and\ \bibinfo {author} {\bibfnamefont {L.-M.}\
  \bibnamefont {Duan}},\ }\bibfield  {title} {\bibinfo {title} {Experimental
  realization of the {Rabi-Hubbard} model with trapped ions},\ }\href
  {https://doi.org/10.1103/PhysRevLett.128.160504} {\bibfield  {journal}
  {\bibinfo  {journal} {Phys. Rev. Lett.}\ }\textbf {\bibinfo {volume} {128}},\
  \bibinfo {pages} {160504} (\bibinfo {year} {2022})}\BibitemShut {NoStop}%
\bibitem [{\citenamefont {Zhao}\ \emph
  {et~al.}(2017{\natexlab{a}})\citenamefont {Zhao}, \citenamefont {Tan},
  \citenamefont {Yu}, \citenamefont {Zhu},\ and\ \citenamefont
  {Yu}}]{PhysRevA.95.063849}%
  \BibitemOpen
  \bibfield  {author} {\bibinfo {author} {\bibfnamefont {P.}~\bibnamefont
  {Zhao}}, \bibinfo {author} {\bibfnamefont {X.}~\bibnamefont {Tan}}, \bibinfo
  {author} {\bibfnamefont {H.}~\bibnamefont {Yu}}, \bibinfo {author}
  {\bibfnamefont {S.-L.}\ \bibnamefont {Zhu}},\ and\ \bibinfo {author}
  {\bibfnamefont {Y.}~\bibnamefont {Yu}},\ }\bibfield  {title} {\bibinfo
  {title} {Simultaneously exciting two atoms with photon-mediated {Raman}
  interactions},\ }\href {https://doi.org/10.1103/PhysRevA.95.063848}
  {\bibfield  {journal} {\bibinfo  {journal} {Phys. Rev. A}\ }\textbf {\bibinfo
  {volume} {95}},\ \bibinfo {pages} {063848} (\bibinfo {year}
  {2017}{\natexlab{a}})}\BibitemShut {NoStop}%
\bibitem [{\citenamefont {Kockum}\ \emph {et~al.}(2019)\citenamefont {Kockum},
  \citenamefont {Miranowicz}, \citenamefont {De~Liberato}, \citenamefont
  {Savasta},\ and\ \citenamefont {Nori}}]{4WOS:000542185800009}%
  \BibitemOpen
  \bibfield  {author} {\bibinfo {author} {\bibfnamefont {A.~F.}\ \bibnamefont
  {Kockum}}, \bibinfo {author} {\bibfnamefont {A.}~\bibnamefont {Miranowicz}},
  \bibinfo {author} {\bibfnamefont {S.}~\bibnamefont {De~Liberato}}, \bibinfo
  {author} {\bibfnamefont {S.}~\bibnamefont {Savasta}},\ and\ \bibinfo {author}
  {\bibfnamefont {F.}~\bibnamefont {Nori}},\ }\bibfield  {title} {\bibinfo
  {title} {Ultrastrong coupling between light and matter},\ }\href
  {https://doi.org/10.1038/s42254-018-0006-2} {\bibfield  {journal} {\bibinfo
  {journal} {Nat. Rev. Phys.}\ }\textbf {\bibinfo {volume} {1}},\ \bibinfo
  {pages} {19–40} (\bibinfo {year} {2019})}\BibitemShut {NoStop}%
\bibitem [{\citenamefont {Forn-D\'{\i}az}\ \emph
  {et~al.}(2019{\natexlab{a}})\citenamefont {Forn-D\'{\i}az}, \citenamefont
  {Lamata}, \citenamefont {Rico}, \citenamefont {Kono},\ and\ \citenamefont
  {Solano}}]{5WOS:000470901400001}%
  \BibitemOpen
  \bibfield  {author} {\bibinfo {author} {\bibfnamefont {P.}~\bibnamefont
  {Forn-D\'{\i}az}}, \bibinfo {author} {\bibfnamefont {L.}~\bibnamefont
  {Lamata}}, \bibinfo {author} {\bibfnamefont {E.}~\bibnamefont {Rico}},
  \bibinfo {author} {\bibfnamefont {J.}~\bibnamefont {Kono}},\ and\ \bibinfo
  {author} {\bibfnamefont {E.}~\bibnamefont {Solano}},\ }\bibfield  {title}
  {\bibinfo {title} {Ultrastrong coupling regimes of light-matter
  interaction},\ }\href {https://doi.org/10.1103/RevModPhys.91.025005}
  {\bibfield  {journal} {\bibinfo  {journal} {Rev. Mod. Phys.}\ }\textbf
  {\bibinfo {volume} {91}},\ \bibinfo {pages} {025005} (\bibinfo {year}
  {2019}{\natexlab{a}})}\BibitemShut {NoStop}%
\bibitem [{\citenamefont {Macr\`{\i}}\ \emph {et~al.}(2022)\citenamefont
  {Macr\`{\i}}, \citenamefont {Minganti}, \citenamefont {Kockum}, \citenamefont
  {Ridolfo}, \citenamefont {Savasta},\ and\ \citenamefont
  {Nori}}]{PhysRevA.105.023720}%
  \BibitemOpen
  \bibfield  {author} {\bibinfo {author} {\bibfnamefont {V.}~\bibnamefont
  {Macr\`{\i}}}, \bibinfo {author} {\bibfnamefont {F.}~\bibnamefont
  {Minganti}}, \bibinfo {author} {\bibfnamefont {A.~F.}\ \bibnamefont
  {Kockum}}, \bibinfo {author} {\bibfnamefont {A.}~\bibnamefont {Ridolfo}},
  \bibinfo {author} {\bibfnamefont {S.}~\bibnamefont {Savasta}},\ and\ \bibinfo
  {author} {\bibfnamefont {F.}~\bibnamefont {Nori}},\ }\bibfield  {title}
  {\bibinfo {title} {Revealing higher-order light and matter energy exchanges
  using quantum trajectories in ultrastrong coupling},\ }\href
  {https://doi.org/10.1103/PhysRevA.105.023720} {\bibfield  {journal} {\bibinfo
   {journal} {Phys. Rev. A}\ }\textbf {\bibinfo {volume} {105}},\ \bibinfo
  {pages} {023720} (\bibinfo {year} {2022})}\BibitemShut {NoStop}%
\bibitem [{\citenamefont {S\'anchez Mu\~noz}\ \emph {et~al.}(2020)\citenamefont
  {S\'anchez Mu\~noz}, \citenamefont {Kockum}, \citenamefont {Miranowicz},\
  and\ \citenamefont {Nori}}]{PhysRevA.102.033716}%
  \BibitemOpen
  \bibfield  {author} {\bibinfo {author} {\bibfnamefont {C.}~\bibnamefont
  {S\'anchez Mu\~noz}}, \bibinfo {author} {\bibfnamefont {A.~F.}\ \bibnamefont
  {Kockum}}, \bibinfo {author} {\bibfnamefont {A.}~\bibnamefont {Miranowicz}},\
  and\ \bibinfo {author} {\bibfnamefont {F.}~\bibnamefont {Nori}},\ }\bibfield
  {title} {\bibinfo {title} {Simulating ultrastrong-coupling processes breaking
  parity conservation in jaynes-cummings systems},\ }\href
  {https://doi.org/10.1103/PhysRevA.102.033716} {\bibfield  {journal} {\bibinfo
   {journal} {Phys. Rev. A}\ }\textbf {\bibinfo {volume} {102}},\ \bibinfo
  {pages} {033716} (\bibinfo {year} {2020})}\BibitemShut {NoStop}%
\bibitem [{\citenamefont {Di~Stefano}\ \emph {et~al.}(2019)\citenamefont
  {Di~Stefano}, \citenamefont {Settineri}, \citenamefont {Macr\`{\i}},
  \citenamefont {Ridolfo}, \citenamefont {Stassi}, \citenamefont {Kockum},
  \citenamefont {Savasta},\ and\ \citenamefont
  {Nori}}]{PhysRevLett.122.030402}%
  \BibitemOpen
  \bibfield  {author} {\bibinfo {author} {\bibfnamefont {O.}~\bibnamefont
  {Di~Stefano}}, \bibinfo {author} {\bibfnamefont {A.}~\bibnamefont
  {Settineri}}, \bibinfo {author} {\bibfnamefont {V.}~\bibnamefont
  {Macr\`{\i}}}, \bibinfo {author} {\bibfnamefont {A.}~\bibnamefont {Ridolfo}},
  \bibinfo {author} {\bibfnamefont {R.}~\bibnamefont {Stassi}}, \bibinfo
  {author} {\bibfnamefont {A.~F.}\ \bibnamefont {Kockum}}, \bibinfo {author}
  {\bibfnamefont {S.}~\bibnamefont {Savasta}},\ and\ \bibinfo {author}
  {\bibfnamefont {F.}~\bibnamefont {Nori}},\ }\bibfield  {title} {\bibinfo
  {title} {Interaction of mechanical oscillators mediated by the exchange of
  virtual photon pairs},\ }\href
  {https://doi.org/10.1103/PhysRevLett.122.030402} {\bibfield  {journal}
  {\bibinfo  {journal} {Phys. Rev. Lett.}\ }\textbf {\bibinfo {volume} {122}},\
  \bibinfo {pages} {030402} (\bibinfo {year} {2019})}\BibitemShut {NoStop}%
\bibitem [{\citenamefont {Kockum}\ \emph {et~al.}(2017)\citenamefont {Kockum},
  \citenamefont {Miranowicz}, \citenamefont {Macr\`{\i}}, \citenamefont
  {Savasta},\ and\ \citenamefont {Nori}}]{-PhysRevA.95.063849}%
  \BibitemOpen
  \bibfield  {author} {\bibinfo {author} {\bibfnamefont {A.~F.}\ \bibnamefont
  {Kockum}}, \bibinfo {author} {\bibfnamefont {A.}~\bibnamefont {Miranowicz}},
  \bibinfo {author} {\bibfnamefont {V.}~\bibnamefont {Macr\`{\i}}}, \bibinfo
  {author} {\bibfnamefont {S.}~\bibnamefont {Savasta}},\ and\ \bibinfo {author}
  {\bibfnamefont {F.}~\bibnamefont {Nori}},\ }\bibfield  {title} {\bibinfo
  {title} {Deterministic quantum nonlinear optics with single atoms and virtual
  photons},\ }\href {https://doi.org/10.1103/PhysRevA.95.063849} {\bibfield
  {journal} {\bibinfo  {journal} {Phys. Rev. A}\ }\textbf {\bibinfo {volume}
  {95}},\ \bibinfo {pages} {063849} (\bibinfo {year} {2017})}\BibitemShut
  {NoStop}%
\bibitem [{\citenamefont {Chen}\ \emph
  {et~al.}(2024{\natexlab{b}})\citenamefont {Chen}, \citenamefont {Peng},
  \citenamefont {Chen}, \citenamefont {Liu}, \citenamefont {Chen},
  \citenamefont {Zhu},\ and\ \citenamefont {Lu}}]{Chen2024}%
  \BibitemOpen
  \bibfield  {author} {\bibinfo {author} {\bibfnamefont {Z.-X.}\ \bibnamefont
  {Chen}}, \bibinfo {author} {\bibfnamefont {Y.-G.}\ \bibnamefont {Peng}},
  \bibinfo {author} {\bibfnamefont {Z.-G.}\ \bibnamefont {Chen}}, \bibinfo
  {author} {\bibfnamefont {Y.}~\bibnamefont {Liu}}, \bibinfo {author}
  {\bibfnamefont {P.}~\bibnamefont {Chen}}, \bibinfo {author} {\bibfnamefont
  {X.-F.}\ \bibnamefont {Zhu}},\ and\ \bibinfo {author} {\bibfnamefont {Y.-Q.}\
  \bibnamefont {Lu}},\ }\bibfield  {title} {\bibinfo {title} {Robust temporal
  adiabatic passage with perfect frequency conversion between detuned acoustic
  cavities},\ }\href {https://doi.org/10.1038/s41467-024-45932-6} {\bibfield
  {journal} {\bibinfo  {journal} {Nat. Commun.}\ }\textbf {\bibinfo {volume}
  {15}},\ \bibinfo {pages} {1478} (\bibinfo {year}
  {2024}{\natexlab{b}})}\BibitemShut {NoStop}%
\bibitem [{\citenamefont {Corona}\ \emph {et~al.}(2011)\citenamefont {Corona},
  \citenamefont {Garay-Palmett},\ and\ \citenamefont
  {U'Ren}}]{PhysRevA.84.033823}%
  \BibitemOpen
  \bibfield  {author} {\bibinfo {author} {\bibfnamefont {M.}~\bibnamefont
  {Corona}}, \bibinfo {author} {\bibfnamefont {K.}~\bibnamefont
  {Garay-Palmett}},\ and\ \bibinfo {author} {\bibfnamefont {A.~B.}\
  \bibnamefont {U'Ren}},\ }\bibfield  {title} {\bibinfo {title} {Third-order
  spontaneous parametric down-conversion in thin optical fibers as a
  photon-triplet source},\ }\href {https://doi.org/10.1103/PhysRevA.84.033823}
  {\bibfield  {journal} {\bibinfo  {journal} {Phys. Rev. A}\ }\textbf {\bibinfo
  {volume} {84}},\ \bibinfo {pages} {033823} (\bibinfo {year}
  {2011})}\BibitemShut {NoStop}%
\bibitem [{\citenamefont {Dot}\ \emph {et~al.}(2012)\citenamefont {Dot},
  \citenamefont {Borne}, \citenamefont {Boulanger}, \citenamefont {Bencheikh},\
  and\ \citenamefont {Levenson}}]{PhysRevA.85.023809}%
  \BibitemOpen
  \bibfield  {author} {\bibinfo {author} {\bibfnamefont {A.}~\bibnamefont
  {Dot}}, \bibinfo {author} {\bibfnamefont {A.}~\bibnamefont {Borne}}, \bibinfo
  {author} {\bibfnamefont {B.}~\bibnamefont {Boulanger}}, \bibinfo {author}
  {\bibfnamefont {K.}~\bibnamefont {Bencheikh}},\ and\ \bibinfo {author}
  {\bibfnamefont {J.~A.}\ \bibnamefont {Levenson}},\ }\bibfield  {title}
  {\bibinfo {title} {Quantum theory analysis of triple photons generated by a
  ${\ensuremath{\chi}}^{(3)}$ process},\ }\href
  {https://doi.org/10.1103/PhysRevA.85.023809} {\bibfield  {journal} {\bibinfo
  {journal} {Phys. Rev. A}\ }\textbf {\bibinfo {volume} {85}},\ \bibinfo
  {pages} {023809} (\bibinfo {year} {2012})}\BibitemShut {NoStop}%
\bibitem [{\citenamefont {Akbari}\ \emph {et~al.}(2016)\citenamefont {Akbari},
  \citenamefont {Andrianov},\ and\ \citenamefont {Kalachev}}]{Akbari2016}%
  \BibitemOpen
  \bibfield  {author} {\bibinfo {author} {\bibfnamefont {M.}~\bibnamefont
  {Akbari}}, \bibinfo {author} {\bibfnamefont {S.~N.}\ \bibnamefont
  {Andrianov}},\ and\ \bibinfo {author} {\bibfnamefont {A.~A.}\ \bibnamefont
  {Kalachev}},\ }\bibfield  {title} {\bibinfo {title} {Single-photon frequency
  conversion via interaction with a three-level atom coupled to a microdisk},\
  }\href {https://doi.org/10.1088/1555-6611/27/2/025202} {\bibfield  {journal}
  {\bibinfo  {journal} {Laser Phys.}\ }\textbf {\bibinfo {volume} {27}},\
  \bibinfo {pages} {025202} (\bibinfo {year} {2016})}\BibitemShut {NoStop}%
\bibitem [{\citenamefont {Longhi}(2022)}]{PhysRevA.106.053503}%
  \BibitemOpen
  \bibfield  {author} {\bibinfo {author} {\bibfnamefont {S.}~\bibnamefont
  {Longhi}},\ }\bibfield  {title} {\bibinfo {title} {Topological aspects in
  nonlinear optical frequency conversion},\ }\href
  {https://doi.org/10.1103/PhysRevA.106.053503} {\bibfield  {journal} {\bibinfo
   {journal} {Phys. Rev. A}\ }\textbf {\bibinfo {volume} {106}},\ \bibinfo
  {pages} {053503} (\bibinfo {year} {2022})}\BibitemShut {NoStop}%
\bibitem [{\citenamefont {Devaux}\ and\ \citenamefont
  {Lantz}(2000)}]{PhysRevLett.85.2308}%
  \BibitemOpen
  \bibfield  {author} {\bibinfo {author} {\bibfnamefont {F.}~\bibnamefont
  {Devaux}}\ and\ \bibinfo {author} {\bibfnamefont {E.}~\bibnamefont {Lantz}},\
  }\bibfield  {title} {\bibinfo {title} {Gain in phase sensitive parametric
  image amplification},\ }\href {https://doi.org/10.1103/PhysRevLett.85.2308}
  {\bibfield  {journal} {\bibinfo  {journal} {Phys. Rev. Lett.}\ }\textbf
  {\bibinfo {volume} {85}},\ \bibinfo {pages} {2308} (\bibinfo {year}
  {2000})}\BibitemShut {NoStop}%
\bibitem [{\citenamefont {Alam}\ \emph {et~al.}(2017)\citenamefont {Alam},
  \citenamefont {Beerepoot},\ and\ \citenamefont {Ruud}}]{Alam2017}%
  \BibitemOpen
  \bibfield  {author} {\bibinfo {author} {\bibfnamefont {M.}~\bibnamefont
  {Alam}}, \bibinfo {author} {\bibfnamefont {M.}~\bibnamefont {Beerepoot}},\
  and\ \bibinfo {author} {\bibfnamefont {K.}~\bibnamefont {Ruud}},\ }\bibfield
  {title} {\bibinfo {title} {Channel interference in multiphoton absorption},\
  }\href {https://doi.org/10.1063/1.4990438} {\bibfield  {journal} {\bibinfo
  {journal} {J. Chem. Phys.}\ }\textbf {\bibinfo {volume} {146}},\ \bibinfo
  {pages} {244116} (\bibinfo {year} {2017})}\BibitemShut {NoStop}%
\bibitem [{\citenamefont {Caumes}\ \emph {et~al.}(2002)\citenamefont {Caumes},
  \citenamefont {Videau}, \citenamefont {Rouyer},\ and\ \citenamefont
  {Freysz}}]{PhysRevLett.89.047401}%
  \BibitemOpen
  \bibfield  {author} {\bibinfo {author} {\bibfnamefont {J.~P.}\ \bibnamefont
  {Caumes}}, \bibinfo {author} {\bibfnamefont {L.}~\bibnamefont {Videau}},
  \bibinfo {author} {\bibfnamefont {C.}~\bibnamefont {Rouyer}},\ and\ \bibinfo
  {author} {\bibfnamefont {E.}~\bibnamefont {Freysz}},\ }\bibfield  {title}
  {\bibinfo {title} {Kerr-like nonlinearity induced via terahertz generation
  and the electro-optical effect in zinc blende crystals},\ }\href
  {https://doi.org/10.1103/PhysRevLett.89.047401} {\bibfield  {journal}
  {\bibinfo  {journal} {Phys. Rev. Lett.}\ }\textbf {\bibinfo {volume} {89}},\
  \bibinfo {pages} {047401} (\bibinfo {year} {2002})}\BibitemShut {NoStop}%
\bibitem [{\citenamefont {Garziano}\ \emph {et~al.}(2015)\citenamefont
  {Garziano}, \citenamefont {Stassi}, \citenamefont {Macr\`{\i}}, \citenamefont
  {Kockum}, \citenamefont {Savasta},\ and\ \citenamefont
  {Nori}}]{8WOS:000366730900005}%
  \BibitemOpen
  \bibfield  {author} {\bibinfo {author} {\bibfnamefont {L.}~\bibnamefont
  {Garziano}}, \bibinfo {author} {\bibfnamefont {R.}~\bibnamefont {Stassi}},
  \bibinfo {author} {\bibfnamefont {V.}~\bibnamefont {Macr\`{\i}}}, \bibinfo
  {author} {\bibfnamefont {A.~F.}\ \bibnamefont {Kockum}}, \bibinfo {author}
  {\bibfnamefont {S.}~\bibnamefont {Savasta}},\ and\ \bibinfo {author}
  {\bibfnamefont {F.}~\bibnamefont {Nori}},\ }\bibfield  {title} {\bibinfo
  {title} {Multiphoton quantum {Rabi} oscillations in ultrastrong cavity qed},\
  }\href {https://doi.org/10.1103/PhysRevA.92.063830} {\bibfield  {journal}
  {\bibinfo  {journal} {Phys. Rev. A}\ }\textbf {\bibinfo {volume} {92}},\
  \bibinfo {pages} {063830} (\bibinfo {year} {2015})}\BibitemShut {NoStop}%
\bibitem [{\citenamefont {Garziano}\ \emph {et~al.}(2016)\citenamefont
  {Garziano}, \citenamefont {Macr\`{\i}}, \citenamefont {Stassi}, \citenamefont
  {Di~Stefano}, \citenamefont {Nori},\ and\ \citenamefont
  {Savasta}}]{PhysRevLett.117.043601}%
  \BibitemOpen
  \bibfield  {author} {\bibinfo {author} {\bibfnamefont {L.}~\bibnamefont
  {Garziano}}, \bibinfo {author} {\bibfnamefont {V.}~\bibnamefont
  {Macr\`{\i}}}, \bibinfo {author} {\bibfnamefont {R.}~\bibnamefont {Stassi}},
  \bibinfo {author} {\bibfnamefont {O.}~\bibnamefont {Di~Stefano}}, \bibinfo
  {author} {\bibfnamefont {F.}~\bibnamefont {Nori}},\ and\ \bibinfo {author}
  {\bibfnamefont {S.}~\bibnamefont {Savasta}},\ }\bibfield  {title} {\bibinfo
  {title} {One photon can simultaneously excite two or more atoms},\ }\href
  {https://doi.org/10.1103/PhysRevLett.117.043601} {\bibfield  {journal}
  {\bibinfo  {journal} {Phys. Rev. Lett.}\ }\textbf {\bibinfo {volume} {117}},\
  \bibinfo {pages} {043601} (\bibinfo {year} {2016})}\BibitemShut {NoStop}%
\bibitem [{\citenamefont {Zheng}\ \emph {et~al.}(2013)\citenamefont {Zheng},
  \citenamefont {Saldanha}, \citenamefont {Rios~Leite},\ and\ \citenamefont
  {Fabre}}]{PhysRevA.88.033822}%
  \BibitemOpen
  \bibfield  {author} {\bibinfo {author} {\bibfnamefont {Z.}~\bibnamefont
  {Zheng}}, \bibinfo {author} {\bibfnamefont {P.~L.}\ \bibnamefont {Saldanha}},
  \bibinfo {author} {\bibfnamefont {J.~R.}\ \bibnamefont {Rios~Leite}},\ and\
  \bibinfo {author} {\bibfnamefont {C.}~\bibnamefont {Fabre}},\ }\bibfield
  {title} {\bibinfo {title} {Two-photon--two-atom excitation by correlated
  light states},\ }\href {https://doi.org/10.1103/PhysRevA.88.033822}
  {\bibfield  {journal} {\bibinfo  {journal} {Phys. Rev. A}\ }\textbf {\bibinfo
  {volume} {88}},\ \bibinfo {pages} {033822} (\bibinfo {year}
  {2013})}\BibitemShut {NoStop}%
\bibitem [{\citenamefont {Ridolfo}\ \emph
  {et~al.}(2012{\natexlab{a}})\citenamefont {Ridolfo}, \citenamefont {Leib},
  \citenamefont {Savasta},\ and\ \citenamefont {Hartmann}}]{5Ridolfo2012}%
  \BibitemOpen
  \bibfield  {author} {\bibinfo {author} {\bibfnamefont {A.}~\bibnamefont
  {Ridolfo}}, \bibinfo {author} {\bibfnamefont {M.}~\bibnamefont {Leib}},
  \bibinfo {author} {\bibfnamefont {S.}~\bibnamefont {Savasta}},\ and\ \bibinfo
  {author} {\bibfnamefont {M.~J.}\ \bibnamefont {Hartmann}},\ }\bibfield
  {title} {\bibinfo {title} {Photon blockade in the ultrastrong coupling
  regime},\ }\href {https://doi.org/10.1103/PhysRevLett.109.193602} {\bibfield
  {journal} {\bibinfo  {journal} {Phys. Rev. Lett.}\ }\textbf {\bibinfo
  {volume} {109}},\ \bibinfo {pages} {193602} (\bibinfo {year}
  {2012}{\natexlab{a}})}\BibitemShut {NoStop}%
\bibitem [{\citenamefont {Felicetti}\ \emph {et~al.}(2015)\citenamefont
  {Felicetti}, \citenamefont {Douce}, \citenamefont {Romero}, \citenamefont
  {Milman},\ and\ \citenamefont {Solano}}]{5Felicetti2015}%
  \BibitemOpen
  \bibfield  {author} {\bibinfo {author} {\bibfnamefont {S.}~\bibnamefont
  {Felicetti}}, \bibinfo {author} {\bibfnamefont {T.}~\bibnamefont {Douce}},
  \bibinfo {author} {\bibfnamefont {G.}~\bibnamefont {Romero}}, \bibinfo
  {author} {\bibfnamefont {P.}~\bibnamefont {Milman}},\ and\ \bibinfo {author}
  {\bibfnamefont {E.}~\bibnamefont {Solano}},\ }\bibfield  {title} {\bibinfo
  {title} {Parity-dependent {state} {engineering} and {tomography} in the
  ultrastrong coupling regime},\ }\href {https://doi.org/10.1038/srep11818}
  {\bibfield  {journal} {\bibinfo  {journal} {Sci. Rep.}\ }\textbf {\bibinfo
  {volume} {5}},\ \bibinfo {pages} {11818} (\bibinfo {year}
  {2015})}\BibitemShut {NoStop}%
\bibitem [{\citenamefont {Cao}\ \emph {et~al.}(2011)\citenamefont {Cao},
  \citenamefont {You}, \citenamefont {Zheng},\ and\ \citenamefont
  {Nori}}]{5Cao2011}%
  \BibitemOpen
  \bibfield  {author} {\bibinfo {author} {\bibfnamefont {X.}~\bibnamefont
  {Cao}}, \bibinfo {author} {\bibfnamefont {J.~Q.}\ \bibnamefont {You}},
  \bibinfo {author} {\bibfnamefont {H.}~\bibnamefont {Zheng}},\ and\ \bibinfo
  {author} {\bibfnamefont {F.}~\bibnamefont {Nori}},\ }\bibfield  {title}
  {\bibinfo {title} {A qubit strongly coupled to a resonant cavity: asymmetry
  of the spontaneous emission spectrum beyond the rotating wave
  approximation},\ }\href {https://doi.org/10.1088/1367-2630/13/7/073002}
  {\bibfield  {journal} {\bibinfo  {journal} {New J. Phys}\ }\textbf {\bibinfo
  {volume} {13}},\ \bibinfo {pages} {073002} (\bibinfo {year}
  {2011})}\BibitemShut {NoStop}%
\bibitem [{\citenamefont {Forn-D\'{\i}az}\ \emph {et~al.}(2010)\citenamefont
  {Forn-D\'{\i}az}, \citenamefont {Lisenfeld}, \citenamefont {Marcos},
  \citenamefont {Garc\'{\i}a-Ripoll}, \citenamefont {Solano}, \citenamefont
  {Harmans},\ and\ \citenamefont {Mooij}}]{6WOS:000286732400017}%
  \BibitemOpen
  \bibfield  {author} {\bibinfo {author} {\bibfnamefont {P.}~\bibnamefont
  {Forn-D\'{\i}az}}, \bibinfo {author} {\bibfnamefont {J.}~\bibnamefont
  {Lisenfeld}}, \bibinfo {author} {\bibfnamefont {D.}~\bibnamefont {Marcos}},
  \bibinfo {author} {\bibfnamefont {J.~J.}\ \bibnamefont {Garc\'{\i}a-Ripoll}},
  \bibinfo {author} {\bibfnamefont {E.}~\bibnamefont {Solano}}, \bibinfo
  {author} {\bibfnamefont {C.~J. P.~M.}\ \bibnamefont {Harmans}},\ and\
  \bibinfo {author} {\bibfnamefont {J.~E.}\ \bibnamefont {Mooij}},\ }\bibfield
  {title} {\bibinfo {title} {Observation of the {Bloch-Siegert} shift in a
  qubit-oscillator system in the ultrastrong coupling regime},\ }\href
  {https://doi.org/10.1103/PhysRevLett.105.237001} {\bibfield  {journal}
  {\bibinfo  {journal} {Phys. Rev. Lett.}\ }\textbf {\bibinfo {volume} {105}},\
  \bibinfo {pages} {237001} (\bibinfo {year} {2010})}\BibitemShut {NoStop}%
\bibitem [{\citenamefont {Zhao}\ \emph
  {et~al.}(2017{\natexlab{b}})\citenamefont {Zhao}, \citenamefont {Tan},
  \citenamefont {Yu}, \citenamefont {Zhu},\ and\ \citenamefont
  {Yu}}]{13Garziano2016}%
  \BibitemOpen
  \bibfield  {author} {\bibinfo {author} {\bibfnamefont {P.}~\bibnamefont
  {Zhao}}, \bibinfo {author} {\bibfnamefont {X.-S.}\ \bibnamefont {Tan}},
  \bibinfo {author} {\bibfnamefont {H.-F.}\ \bibnamefont {Yu}}, \bibinfo
  {author} {\bibfnamefont {S.-L.}\ \bibnamefont {Zhu}},\ and\ \bibinfo {author}
  {\bibfnamefont {Y.}~\bibnamefont {Yu}},\ }\bibfield  {title} {\bibinfo
  {title} {Simultaneously exciting two atoms with photon-mediated {Raman}
  interactions},\ }\href {https://doi.org/10.1103/PhysRevA.95.063848}
  {\bibfield  {journal} {\bibinfo  {journal} {Phys. Rev. A}\ }\textbf {\bibinfo
  {volume} {95}},\ \bibinfo {pages} {063848} (\bibinfo {year}
  {2017}{\natexlab{b}})}\BibitemShut {NoStop}%
\bibitem [{\citenamefont {Ma}\ and\ \citenamefont {Law}(2015)}]{16Ma2015}%
  \BibitemOpen
  \bibfield  {author} {\bibinfo {author} {\bibfnamefont {K.~K.~W.}\
  \bibnamefont {Ma}}\ and\ \bibinfo {author} {\bibfnamefont {C.~K.}\
  \bibnamefont {Law}},\ }\bibfield  {title} {\bibinfo {title} {Three-photon
  resonance and adiabatic passage in the large-detuning {Rabi} model},\ }\href
  {https://doi.org/10.1103/PhysRevA.92.023842} {\bibfield  {journal} {\bibinfo
  {journal} {Phys. Rev. A}\ }\textbf {\bibinfo {volume} {92}},\ \bibinfo
  {pages} {023842} (\bibinfo {year} {2015})}\BibitemShut {NoStop}%
\bibitem [{\citenamefont {Felicetti}\ \emph {et~al.}(2018)\citenamefont
  {Felicetti}, \citenamefont {Rossatto}, \citenamefont {Rico}, \citenamefont
  {Solano},\ and\ \citenamefont {Forn-D\'{\i}az}}]{15Felicetti2018}%
  \BibitemOpen
  \bibfield  {author} {\bibinfo {author} {\bibfnamefont {S.}~\bibnamefont
  {Felicetti}}, \bibinfo {author} {\bibfnamefont {D.~Z.}\ \bibnamefont
  {Rossatto}}, \bibinfo {author} {\bibfnamefont {E.}~\bibnamefont {Rico}},
  \bibinfo {author} {\bibfnamefont {E.}~\bibnamefont {Solano}},\ and\ \bibinfo
  {author} {\bibfnamefont {P.}~\bibnamefont {Forn-D\'{\i}az}},\ }\bibfield
  {title} {\bibinfo {title} {Two-photon quantum {Rabi} model with
  superconducting circuits},\ }\href
  {https://doi.org/10.1103/PhysRevA.97.013851} {\bibfield  {journal} {\bibinfo
  {journal} {Phys. Rev. A}\ }\textbf {\bibinfo {volume} {97}},\ \bibinfo
  {pages} {013851} (\bibinfo {year} {2018})}\BibitemShut {NoStop}%
\bibitem [{\citenamefont {Liu}\ and\ \citenamefont
  {Huang}(2024)}]{10WOS:001112616000002}%
  \BibitemOpen
  \bibfield  {author} {\bibinfo {author} {\bibfnamefont {C.}~\bibnamefont
  {Liu}}\ and\ \bibinfo {author} {\bibfnamefont {J.-F.}\ \bibnamefont
  {Huang}},\ }\bibfield  {title} {\bibinfo {title} {Quantum phase transition of
  the {Jaynes}-{Cummings} model},\ }\href
  {https://doi.org/10.1007/s11433-023-2243-7} {\bibfield  {journal} {\bibinfo
  {journal} {Sci. China Phys. Mech. Astron.}\ }\textbf {\bibinfo {volume}
  {67}},\ \bibinfo {pages} {210311} (\bibinfo {year} {2024})}\BibitemShut
  {NoStop}%
\bibitem [{\citenamefont {Bertet}\ \emph {et~al.}(2002)\citenamefont {Bertet},
  \citenamefont {Osnaghi}, \citenamefont {Milman}, \citenamefont {Auffeves},
  \citenamefont {Maioli}, \citenamefont {Brune}, \citenamefont {Raimond},\ and\
  \citenamefont {Haroche}}]{PhysRevLett.88.143601}%
  \BibitemOpen
  \bibfield  {author} {\bibinfo {author} {\bibfnamefont {P.}~\bibnamefont
  {Bertet}}, \bibinfo {author} {\bibfnamefont {S.}~\bibnamefont {Osnaghi}},
  \bibinfo {author} {\bibfnamefont {P.}~\bibnamefont {Milman}}, \bibinfo
  {author} {\bibfnamefont {A.}~\bibnamefont {Auffeves}}, \bibinfo {author}
  {\bibfnamefont {P.}~\bibnamefont {Maioli}}, \bibinfo {author} {\bibfnamefont
  {M.}~\bibnamefont {Brune}}, \bibinfo {author} {\bibfnamefont {J.~M.}\
  \bibnamefont {Raimond}},\ and\ \bibinfo {author} {\bibfnamefont
  {S.}~\bibnamefont {Haroche}},\ }\bibfield  {title} {\bibinfo {title}
  {Generating and probing a two-photon {Fock} state with a single atom in a
  cavity},\ }\href {https://doi.org/10.1103/PhysRevLett.88.143601} {\bibfield
  {journal} {\bibinfo  {journal} {Phys. Rev. Lett.}\ }\textbf {\bibinfo
  {volume} {88}},\ \bibinfo {pages} {143601} (\bibinfo {year}
  {2002})}\BibitemShut {NoStop}%
\bibitem [{\citenamefont {Shi}\ and\ \citenamefont {Wei}(2014)}]{Shi2014}%
  \BibitemOpen
  \bibfield  {author} {\bibinfo {author} {\bibfnamefont {X.}~\bibnamefont
  {Shi}}\ and\ \bibinfo {author} {\bibfnamefont {L.~F.}\ \bibnamefont {Wei}},\
  }\bibfield  {title} {\bibinfo {title} {High-efficiency single-photon {Fock}
  state production by transitionless quantum driving},\ }\href
  {https://doi.org/10.1088/1612-2011/12/1/015204} {\bibfield  {journal}
  {\bibinfo  {journal} {Laser Phys. Lett.}\ }\textbf {\bibinfo {volume} {12}},\
  \bibinfo {pages} {015204} (\bibinfo {year} {2014})}\BibitemShut {NoStop}%
\bibitem [{\citenamefont {Han}\ \emph {et~al.}(2021)\citenamefont {Han},
  \citenamefont {Wu}, \citenamefont {Wang}, \citenamefont {Xia}, \citenamefont
  {Jiang},\ and\ \citenamefont {Song}}]{PhysRevA.103.032402}%
  \BibitemOpen
  \bibfield  {author} {\bibinfo {author} {\bibfnamefont {J.-X.}\ \bibnamefont
  {Han}}, \bibinfo {author} {\bibfnamefont {J.-L.}\ \bibnamefont {Wu}},
  \bibinfo {author} {\bibfnamefont {Y.}~\bibnamefont {Wang}}, \bibinfo {author}
  {\bibfnamefont {Y.}~\bibnamefont {Xia}}, \bibinfo {author} {\bibfnamefont
  {Y.-Y.}\ \bibnamefont {Jiang}},\ and\ \bibinfo {author} {\bibfnamefont
  {J.}~\bibnamefont {Song}},\ }\bibfield  {title} {\bibinfo {title}
  {Large-scale greenberger-horne-zeilinger states through a topologically
  protected zero-energy mode in a superconducting qutrit-resonator chain},\
  }\href {https://doi.org/10.1103/PhysRevA.103.032402} {\bibfield  {journal}
  {\bibinfo  {journal} {Phys. Rev. A}\ }\textbf {\bibinfo {volume} {103}},\
  \bibinfo {pages} {032402} (\bibinfo {year} {2021})}\BibitemShut {NoStop}%
\bibitem [{\citenamefont {Huang}\ \emph {et~al.}(2017)\citenamefont {Huang},
  \citenamefont {Liao}, \citenamefont {Tian},\ and\ \citenamefont
  {Kuang}}]{17Huang2017}%
  \BibitemOpen
  \bibfield  {author} {\bibinfo {author} {\bibfnamefont {J.-F.}\ \bibnamefont
  {Huang}}, \bibinfo {author} {\bibfnamefont {J.-Q.}\ \bibnamefont {Liao}},
  \bibinfo {author} {\bibfnamefont {L.}~\bibnamefont {Tian}},\ and\ \bibinfo
  {author} {\bibfnamefont {L.-M.}\ \bibnamefont {Kuang}},\ }\bibfield  {title}
  {\bibinfo {title} {Manipulating counter-rotating interactions in the quantum
  {Rabi} model via modulation of the transition frequency of the two-level
  system},\ }\href {https://doi.org/10.1103/PhysRevA.96.043849} {\bibfield
  {journal} {\bibinfo  {journal} {Phys. Rev. A}\ }\textbf {\bibinfo {volume}
  {96}},\ \bibinfo {pages} {043849} (\bibinfo {year} {2017})}\BibitemShut
  {NoStop}%
\bibitem [{\citenamefont {Silveri}\ \emph {et~al.}(2017)\citenamefont
  {Silveri}, \citenamefont {Tuorila}, \citenamefont {Thuneberg},\ and\
  \citenamefont {Paraoanu}}]{18Silveri2017}%
  \BibitemOpen
  \bibfield  {author} {\bibinfo {author} {\bibfnamefont {M.~P.}\ \bibnamefont
  {Silveri}}, \bibinfo {author} {\bibfnamefont {J.~A.}\ \bibnamefont
  {Tuorila}}, \bibinfo {author} {\bibfnamefont {E.~V.}\ \bibnamefont
  {Thuneberg}},\ and\ \bibinfo {author} {\bibfnamefont {G.~S.}\ \bibnamefont
  {Paraoanu}},\ }\bibfield  {title} {\bibinfo {title} {Quantum systems under
  frequency modulation},\ }\href {https://doi.org/10.1088/1361-6633/aa5170}
  {\bibfield  {journal} {\bibinfo  {journal} {Rep. Prog. Phys.}\ }\textbf
  {\bibinfo {volume} {80}},\ \bibinfo {pages} {056002} (\bibinfo {year}
  {2017})}\BibitemShut {NoStop}%
\bibitem [{\citenamefont {Liao}\ \emph {et~al.}(2015)\citenamefont {Liao},
  \citenamefont {Law}, \citenamefont {Kuang},\ and\ \citenamefont
  {Nori}}]{19Liao2015}%
  \BibitemOpen
  \bibfield  {author} {\bibinfo {author} {\bibfnamefont {J.-Q.}\ \bibnamefont
  {Liao}}, \bibinfo {author} {\bibfnamefont {C.~K.}\ \bibnamefont {Law}},
  \bibinfo {author} {\bibfnamefont {L.-M.}\ \bibnamefont {Kuang}},\ and\
  \bibinfo {author} {\bibfnamefont {F.}~\bibnamefont {Nori}},\ }\bibfield
  {title} {\bibinfo {title} {Enhancement of mechanical effects of single
  photons in modulated two-mode optomechanics},\ }\href
  {https://doi.org/10.1103/PhysRevA.92.013822} {\bibfield  {journal} {\bibinfo
  {journal} {Phys. Rev. A}\ }\textbf {\bibinfo {volume} {92}},\ \bibinfo
  {pages} {013822} (\bibinfo {year} {2015})}\BibitemShut {NoStop}%
\bibitem [{\citenamefont {Liao}\ \emph {et~al.}(2016)\citenamefont {Liao},
  \citenamefont {Huang},\ and\ \citenamefont {Tian}}]{20Liao2016}%
  \BibitemOpen
  \bibfield  {author} {\bibinfo {author} {\bibfnamefont {J.-Q.}\ \bibnamefont
  {Liao}}, \bibinfo {author} {\bibfnamefont {J.-F.}\ \bibnamefont {Huang}},\
  and\ \bibinfo {author} {\bibfnamefont {L.}~\bibnamefont {Tian}},\ }\bibfield
  {title} {\bibinfo {title} {Generation of macroscopic schr\"odinger-cat states
  in qubit-oscillator systems},\ }\href
  {https://doi.org/10.1103/PhysRevA.93.033853} {\bibfield  {journal} {\bibinfo
  {journal} {Phys. Rev. A}\ }\textbf {\bibinfo {volume} {93}},\ \bibinfo
  {pages} {033853} (\bibinfo {year} {2016})}\BibitemShut {NoStop}%
\bibitem [{\citenamefont {Beaudoin}\ \emph {et~al.}(2012)\citenamefont
  {Beaudoin}, \citenamefont {Da~Silva}, \citenamefont {Dutton},\ and\
  \citenamefont {Blais}}]{21Beaudoin2012}%
  \BibitemOpen
  \bibfield  {author} {\bibinfo {author} {\bibfnamefont {F.}~\bibnamefont
  {Beaudoin}}, \bibinfo {author} {\bibfnamefont {M.~P.}\ \bibnamefont
  {Da~Silva}}, \bibinfo {author} {\bibfnamefont {Z.}~\bibnamefont {Dutton}},\
  and\ \bibinfo {author} {\bibfnamefont {A.}~\bibnamefont {Blais}},\ }\bibfield
   {title} {\bibinfo {title} {First-order sidebands in circuit {QED} using
  qubit frequency modulation},\ }\href
  {https://doi.org/10.1103/PhysRevA.86.022305} {\bibfield  {journal} {\bibinfo
  {journal} {Phys. Rev. A}\ }\textbf {\bibinfo {volume} {86}},\ \bibinfo
  {pages} {022305} (\bibinfo {year} {2012})}\BibitemShut {NoStop}%
\bibitem [{\citenamefont {Zhang}\ \emph {et~al.}(2024)\citenamefont {Zhang},
  \citenamefont {Wu}, \citenamefont {Sun}, \citenamefont {Wu},\ and\
  \citenamefont {Wang}}]{22Zhang2024}%
  \BibitemOpen
  \bibfield  {author} {\bibinfo {author} {\bibfnamefont {W.}~\bibnamefont
  {Zhang}}, \bibinfo {author} {\bibfnamefont {R.}~\bibnamefont {Wu}}, \bibinfo
  {author} {\bibfnamefont {C.}~\bibnamefont {Sun}}, \bibinfo {author}
  {\bibfnamefont {C.}~\bibnamefont {Wu}},\ and\ \bibinfo {author}
  {\bibfnamefont {G.}~\bibnamefont {Wang}},\ }\bibfield  {title} {\bibinfo
  {title} {Entangling two dicke states in a periodic modulated quantum
  system},\ }\href {https://doi.org/10.1103/PhysRevA.109.013712} {\bibfield
  {journal} {\bibinfo  {journal} {Phys. Rev. A}\ }\textbf {\bibinfo {volume}
  {109}},\ \bibinfo {pages} {013712} (\bibinfo {year} {2024})}\BibitemShut
  {NoStop}%
\bibitem [{\citenamefont {Nourmandipour}\ and\ \citenamefont
  {Mortezapour}(2023)}]{23Nourmandipour2023}%
  \BibitemOpen
  \bibfield  {author} {\bibinfo {author} {\bibfnamefont {A.}~\bibnamefont
  {Nourmandipour}}\ and\ \bibinfo {author} {\bibfnamefont {A.}~\bibnamefont
  {Mortezapour}},\ }\bibfield  {title} {\bibinfo {title} {Frequency–modulated
  qubits in a dissipative cavity: entanglement dynamics and protection},\
  }\href {https://doi.org/10.1007/s11128-023-03992-5} {\bibfield  {journal}
  {\bibinfo  {journal} {Quantum Inf. Process.}\ }\textbf {\bibinfo {volume}
  {22}},\ \bibinfo {pages} {254} (\bibinfo {year} {2023})}\BibitemShut
  {NoStop}%
\bibitem [{\citenamefont {Ashhab}\ \emph {et~al.}(2007)\citenamefont {Ashhab},
  \citenamefont {Johansson}, \citenamefont {Zagoskin},\ and\ \citenamefont
  {Nori}}]{PhysRevA.75.063414}%
  \BibitemOpen
  \bibfield  {author} {\bibinfo {author} {\bibfnamefont {S.}~\bibnamefont
  {Ashhab}}, \bibinfo {author} {\bibfnamefont {J.~R.}\ \bibnamefont
  {Johansson}}, \bibinfo {author} {\bibfnamefont {A.~M.}\ \bibnamefont
  {Zagoskin}},\ and\ \bibinfo {author} {\bibfnamefont {F.}~\bibnamefont
  {Nori}},\ }\bibfield  {title} {\bibinfo {title} {Two-level systems driven by
  large-amplitude fields},\ }\href {https://doi.org/10.1103/PhysRevA.75.063414}
  {\bibfield  {journal} {\bibinfo  {journal} {Phys. Rev. A}\ }\textbf {\bibinfo
  {volume} {75}},\ \bibinfo {pages} {063414} (\bibinfo {year}
  {2007})}\BibitemShut {NoStop}%
\bibitem [{\citenamefont {Zheng}\ \emph {et~al.}(2023)\citenamefont {Zheng},
  \citenamefont {Ning}, \citenamefont {Chen}, \citenamefont {L\"u},
  \citenamefont {Shen}, \citenamefont {Xu}, \citenamefont {Zhang},
  \citenamefont {Xu}, \citenamefont {Li}, \citenamefont {Xia}, \citenamefont
  {Wu}, \citenamefont {Yang}, \citenamefont {Miranowicz}, \citenamefont
  {Lambert}, \citenamefont {Zheng}, \citenamefont {Fan}, \citenamefont {Nori},\
  and\ \citenamefont {Zheng}}]{PhysRevLett.131.113601}%
  \BibitemOpen
  \bibfield  {author} {\bibinfo {author} {\bibfnamefont {R.-H.}\ \bibnamefont
  {Zheng}}, \bibinfo {author} {\bibfnamefont {W.}~\bibnamefont {Ning}},
  \bibinfo {author} {\bibfnamefont {Y.-H.}\ \bibnamefont {Chen}}, \bibinfo
  {author} {\bibfnamefont {J.-H.}\ \bibnamefont {L\"u}}, \bibinfo {author}
  {\bibfnamefont {L.-T.}\ \bibnamefont {Shen}}, \bibinfo {author}
  {\bibfnamefont {K.}~\bibnamefont {Xu}}, \bibinfo {author} {\bibfnamefont
  {Y.-R.}\ \bibnamefont {Zhang}}, \bibinfo {author} {\bibfnamefont
  {D.}~\bibnamefont {Xu}}, \bibinfo {author} {\bibfnamefont {H.}~\bibnamefont
  {Li}}, \bibinfo {author} {\bibfnamefont {Y.}~\bibnamefont {Xia}}, \bibinfo
  {author} {\bibfnamefont {F.}~\bibnamefont {Wu}}, \bibinfo {author}
  {\bibfnamefont {Z.-B.}\ \bibnamefont {Yang}}, \bibinfo {author}
  {\bibfnamefont {A.}~\bibnamefont {Miranowicz}}, \bibinfo {author}
  {\bibfnamefont {N.}~\bibnamefont {Lambert}}, \bibinfo {author} {\bibfnamefont
  {D.}~\bibnamefont {Zheng}}, \bibinfo {author} {\bibfnamefont
  {H.}~\bibnamefont {Fan}}, \bibinfo {author} {\bibfnamefont {F.}~\bibnamefont
  {Nori}},\ and\ \bibinfo {author} {\bibfnamefont {S.-B.}\ \bibnamefont
  {Zheng}},\ }\bibfield  {title} {\bibinfo {title} {Observation of a
  superradiant phase transition with emergent cat states},\ }\href
  {https://doi.org/10.1103/PhysRevLett.131.113601} {\bibfield  {journal}
  {\bibinfo  {journal} {Phys. Rev. Lett.}\ }\textbf {\bibinfo {volume} {131}},\
  \bibinfo {pages} {113601} (\bibinfo {year} {2023})}\BibitemShut {NoStop}%
\bibitem [{\citenamefont {Xie}\ \emph {et~al.}(2014)\citenamefont {Xie},
  \citenamefont {Cui}, \citenamefont {Cao}, \citenamefont {Amico},\ and\
  \citenamefont {Fan}}]{Xie2014}%
  \BibitemOpen
  \bibfield  {author} {\bibinfo {author} {\bibfnamefont {Q.-T.}\ \bibnamefont
  {Xie}}, \bibinfo {author} {\bibfnamefont {S.}~\bibnamefont {Cui}}, \bibinfo
  {author} {\bibfnamefont {J.-P.}\ \bibnamefont {Cao}}, \bibinfo {author}
  {\bibfnamefont {L.}~\bibnamefont {Amico}},\ and\ \bibinfo {author}
  {\bibfnamefont {H.}~\bibnamefont {Fan}},\ }\bibfield  {title} {\bibinfo
  {title} {Anisotropic {Rabi} model},\ }\href
  {https://doi.org/10.1103/PhysRevX.4.021046} {\bibfield  {journal} {\bibinfo
  {journal} {Phys. Rev. X}\ }\textbf {\bibinfo {volume} {4}},\ \bibinfo {pages}
  {021046} (\bibinfo {year} {2014})}\BibitemShut {NoStop}%
\bibitem [{\citenamefont {Zheng}\ and\ \citenamefont
  {Guo}(2000)}]{PhysRevLett.85.2392}%
  \BibitemOpen
  \bibfield  {author} {\bibinfo {author} {\bibfnamefont {S.-B.}\ \bibnamefont
  {Zheng}}\ and\ \bibinfo {author} {\bibfnamefont {G.-C.}\ \bibnamefont
  {Guo}},\ }\bibfield  {title} {\bibinfo {title} {Efficient scheme for two-atom
  entanglement and quantum information processing in cavity qed},\ }\href
  {https://doi.org/10.1103/PhysRevLett.85.2392} {\bibfield  {journal} {\bibinfo
   {journal} {Phys. Rev. Lett.}\ }\textbf {\bibinfo {volume} {85}},\ \bibinfo
  {pages} {2392} (\bibinfo {year} {2000})}\BibitemShut {NoStop}%
\bibitem [{\citenamefont {James}\ and\ \citenamefont
  {Jerke}(2007)}]{James2007}%
  \BibitemOpen
  \bibfield  {author} {\bibinfo {author} {\bibfnamefont {D.~F.}\ \bibnamefont
  {James}}\ and\ \bibinfo {author} {\bibfnamefont {J.}~\bibnamefont {Jerke}},\
  }\bibfield  {title} {\bibinfo {title} {Effective {Hamiltonian} theory and its
  applications in quantum information},\ }\href
  {https://doi.org/10.1139/p07-060} {\bibfield  {journal} {\bibinfo  {journal}
  {Can. J. Phys.}\ }\textbf {\bibinfo {volume} {85}},\ \bibinfo {pages}
  {625–632} (\bibinfo {year} {2007})}\BibitemShut {NoStop}%
\bibitem [{\citenamefont {Shao}\ \emph {et~al.}(2017)\citenamefont {Shao},
  \citenamefont {Wu},\ and\ \citenamefont {Feng}}]{Shao2017}%
  \BibitemOpen
  \bibfield  {author} {\bibinfo {author} {\bibfnamefont {W.}~\bibnamefont
  {Shao}}, \bibinfo {author} {\bibfnamefont {C.}~\bibnamefont {Wu}},\ and\
  \bibinfo {author} {\bibfnamefont {X.-L.}\ \bibnamefont {Feng}},\ }\bibfield
  {title} {\bibinfo {title} {Generalized {James'} effective {Hamiltonian}
  method},\ }\href {https://doi.org/10.1103/PhysRevA.95.032124} {\bibfield
  {journal} {\bibinfo  {journal} {Phys. Rev. A}\ }\textbf {\bibinfo {volume}
  {95}},\ \bibinfo {pages} {032124} (\bibinfo {year} {2017})}\BibitemShut
  {NoStop}%
\bibitem [{\citenamefont {Ridolfo}\ \emph
  {et~al.}(2012{\natexlab{b}})\citenamefont {Ridolfo}, \citenamefont {Leib},
  \citenamefont {Savasta},\ and\ \citenamefont
  {Hartmann}}]{PhysRevLett.109.193602}%
  \BibitemOpen
  \bibfield  {author} {\bibinfo {author} {\bibfnamefont {A.}~\bibnamefont
  {Ridolfo}}, \bibinfo {author} {\bibfnamefont {M.}~\bibnamefont {Leib}},
  \bibinfo {author} {\bibfnamefont {S.}~\bibnamefont {Savasta}},\ and\ \bibinfo
  {author} {\bibfnamefont {M.~J.}\ \bibnamefont {Hartmann}},\ }\bibfield
  {title} {\bibinfo {title} {Photon blockade in the ultrastrong coupling
  regime},\ }\href {https://doi.org/10.1103/PhysRevLett.109.193602} {\bibfield
  {journal} {\bibinfo  {journal} {Phys. Rev. Lett.}\ }\textbf {\bibinfo
  {volume} {109}},\ \bibinfo {pages} {193602} (\bibinfo {year}
  {2012}{\natexlab{b}})}\BibitemShut {NoStop}%
\bibitem [{\citenamefont {Stassi}\ \emph {et~al.}(2013)\citenamefont {Stassi},
  \citenamefont {Ridolfo}, \citenamefont {Di~Stefano}, \citenamefont
  {Hartmann},\ and\ \citenamefont {Savasta}}]{PhysRevLett.110.243601}%
  \BibitemOpen
  \bibfield  {author} {\bibinfo {author} {\bibfnamefont {R.}~\bibnamefont
  {Stassi}}, \bibinfo {author} {\bibfnamefont {A.}~\bibnamefont {Ridolfo}},
  \bibinfo {author} {\bibfnamefont {O.}~\bibnamefont {Di~Stefano}}, \bibinfo
  {author} {\bibfnamefont {M.~J.}\ \bibnamefont {Hartmann}},\ and\ \bibinfo
  {author} {\bibfnamefont {S.}~\bibnamefont {Savasta}},\ }\bibfield  {title}
  {\bibinfo {title} {Spontaneous conversion from virtual to real photons in the
  ultrastrong-coupling regime},\ }\href
  {https://doi.org/10.1103/PhysRevLett.110.243601} {\bibfield  {journal}
  {\bibinfo  {journal} {Phys. Rev. Lett.}\ }\textbf {\bibinfo {volume} {110}},\
  \bibinfo {pages} {243601} (\bibinfo {year} {2013})}\BibitemShut {NoStop}%
\bibitem [{\citenamefont {Huang}\ and\ \citenamefont
  {Law}(2014)}]{PhysRevA.89.033827}%
  \BibitemOpen
  \bibfield  {author} {\bibinfo {author} {\bibfnamefont {J.-F.}\ \bibnamefont
  {Huang}}\ and\ \bibinfo {author} {\bibfnamefont {C.~K.}\ \bibnamefont
  {Law}},\ }\bibfield  {title} {\bibinfo {title} {Photon emission via
  vacuum-dressed intermediate states under ultrastrong coupling},\ }\href
  {https://doi.org/10.1103/PhysRevA.89.033827} {\bibfield  {journal} {\bibinfo
  {journal} {Phys. Rev. A}\ }\textbf {\bibinfo {volume} {89}},\ \bibinfo
  {pages} {033827} (\bibinfo {year} {2014})}\BibitemShut {NoStop}%
\bibitem [{\citenamefont {Crispin~G.}(2004)}]{book1}%
  \BibitemOpen
  \bibfield  {author} {\bibinfo {author} {\bibfnamefont {P.~Z.}\ \bibnamefont
  {Crispin~G.}},\ }\href@noop {} {\emph {\bibinfo {title} {Quantum Noise---A
  Handbook of Markovian and Non-Markovian Quantum Stochastic Methods with
  Applications to Quantum Optics}}}\ (\bibinfo  {publisher} {Springer-Verlag},\
  \bibinfo {year} {2004})\BibitemShut {NoStop}%
\bibitem [{\citenamefont {Li}\ \emph {et~al.}(2013)\citenamefont {Li},
  \citenamefont {Silveri}, \citenamefont {Kumar}, \citenamefont {Pirkkalainen},
  \citenamefont {Vepsäläinen}, \citenamefont {Chien}, \citenamefont
  {Tuorila}, \citenamefont {Sillanpää}, \citenamefont {Hakonen},
  \citenamefont {Thuneberg},\ and\ \citenamefont
  {Paraoanu}}]{li_motional_2013}%
  \BibitemOpen
  \bibfield  {author} {\bibinfo {author} {\bibfnamefont {J.}~\bibnamefont
  {Li}}, \bibinfo {author} {\bibfnamefont {M.~P.}\ \bibnamefont {Silveri}},
  \bibinfo {author} {\bibfnamefont {K.~S.}\ \bibnamefont {Kumar}}, \bibinfo
  {author} {\bibfnamefont {J.~M.}\ \bibnamefont {Pirkkalainen}}, \bibinfo
  {author} {\bibfnamefont {A.}~\bibnamefont {Vepsäläinen}}, \bibinfo {author}
  {\bibfnamefont {W.}~\bibnamefont {Chien}}, \bibinfo {author} {\bibfnamefont
  {J.}~\bibnamefont {Tuorila}}, \bibinfo {author} {\bibfnamefont
  {M.}~\bibnamefont {Sillanpää}}, \bibinfo {author} {\bibfnamefont
  {P.}~\bibnamefont {Hakonen}}, \bibinfo {author} {\bibfnamefont
  {E.}~\bibnamefont {Thuneberg}},\ and\ \bibinfo {author} {\bibfnamefont
  {G.}~\bibnamefont {Paraoanu}},\ }\bibfield  {title} {\bibinfo {title}
  {Motional averaging in a superconducting qubit},\ }\href
  {https://doi.org/10.1038/ncomms2383} {\bibfield  {journal} {\bibinfo
  {journal} {Nat. Commun.}\ }\textbf {\bibinfo {volume} {4}},\ \bibinfo {pages}
  {1420} (\bibinfo {year} {2013})}\BibitemShut {NoStop}%
\bibitem [{\citenamefont {Johansson}\ \emph {et~al.}(2006)\citenamefont
  {Johansson}, \citenamefont {Saito}, \citenamefont {Meno}, \citenamefont
  {Nakano}, \citenamefont {Ueda}, \citenamefont {Semba},\ and\ \citenamefont
  {Takayanagi}}]{PhysRevLett.96.127006}%
  \BibitemOpen
  \bibfield  {author} {\bibinfo {author} {\bibfnamefont {J.}~\bibnamefont
  {Johansson}}, \bibinfo {author} {\bibfnamefont {S.}~\bibnamefont {Saito}},
  \bibinfo {author} {\bibfnamefont {T.}~\bibnamefont {Meno}}, \bibinfo {author}
  {\bibfnamefont {H.}~\bibnamefont {Nakano}}, \bibinfo {author} {\bibfnamefont
  {M.}~\bibnamefont {Ueda}}, \bibinfo {author} {\bibfnamefont {K.}~\bibnamefont
  {Semba}},\ and\ \bibinfo {author} {\bibfnamefont {H.}~\bibnamefont
  {Takayanagi}},\ }\bibfield  {title} {\bibinfo {title} {Vacuum {Rabi}
  oscillations in a macroscopic superconducting qubit {LC} oscillator system},\
  }\href {https://doi.org/10.1103/PhysRevLett.96.127006} {\bibfield  {journal}
  {\bibinfo  {journal} {Phys. Rev. Lett.}\ }\textbf {\bibinfo {volume} {96}},\
  \bibinfo {pages} {127006} (\bibinfo {year} {2006})}\BibitemShut {NoStop}%
\bibitem [{\citenamefont {Forn-D\'{\i}az}\ \emph
  {et~al.}(2019{\natexlab{b}})\citenamefont {Forn-D\'{\i}az}, \citenamefont
  {Lamata}, \citenamefont {Rico}, \citenamefont {Kono},\ and\ \citenamefont
  {Solano}}]{RevModPhys.91.025005}%
  \BibitemOpen
  \bibfield  {author} {\bibinfo {author} {\bibfnamefont {P.}~\bibnamefont
  {Forn-D\'{\i}az}}, \bibinfo {author} {\bibfnamefont {L.}~\bibnamefont
  {Lamata}}, \bibinfo {author} {\bibfnamefont {E.}~\bibnamefont {Rico}},
  \bibinfo {author} {\bibfnamefont {J.}~\bibnamefont {Kono}},\ and\ \bibinfo
  {author} {\bibfnamefont {E.}~\bibnamefont {Solano}},\ }\bibfield  {title}
  {\bibinfo {title} {Ultrastrong coupling regimes of light-matter
  interaction},\ }\href {https://doi.org/10.1103/RevModPhys.91.025005}
  {\bibfield  {journal} {\bibinfo  {journal} {Rev. Mod. Phys.}\ }\textbf
  {\bibinfo {volume} {91}},\ \bibinfo {pages} {025005} (\bibinfo {year}
  {2019}{\natexlab{b}})}\BibitemShut {NoStop}%
\end{thebibliography}%
\end{document}